\documentclass[sigconf,table]{acmart}   

\usepackage{amsmath}
\usepackage{graphicx}
\usepackage{booktabs}
\usepackage{amssymb}
\usepackage{multirow}
\usepackage{subfigure}
\usepackage{algorithm}
\usepackage[noend]{algpseudocode}
\usepackage{xspace}
\usepackage{color}
\usepackage{wasysym}
\usepackage{rotating}   
\usepackage{listings}
\usepackage{balance}
\usepackage{tabu}
\usepackage{pifont}
\usepackage{framed}
\usepackage{enumitem}   
\usepackage[hyphenbreaks]{breakurl}

\usepackage{url}            
\usepackage{hyperref}

\usepackage{fancybox}
\copyrightyear{2021} 
\acmYear{2021} 
\setcopyright{acmcopyright}\acmConference[RAID '21]{24th International Symposium on Research in Attacks, Intrusions and Defenses}{October 6--8, 2021}{San Sebastian, Spain}
\acmBooktitle{24th International Symposium on Research in Attacks, Intrusions and Defenses (RAID '21), October 6--8, 2021, San Sebastian, Spain}
\acmPrice{15.00}
\acmDOI{10.1145/3471621.3471625}
\acmISBN{978-1-4503-9058-3/21/10}

\newcommand{\red}[1]{\textcolor[rgb]{0.00,0.00,0.00}{#1}}

\newcommand{\darkblue}[1]{\textcolor[rgb]{0.00,0.00,0.65}{#1}}
\newcommand{\green}[1]{\textcolor[rgb]{0.00,0.60,0.00}{#1}}

\newcommand{\reb}[1]{\textcolor[RGB]{255,0,0}{#1}}

\newcommand{\cha}{\reb{\ding{55}}}
\newcommand{\gou}{\green{\ding{52}}}
\newcommand{\ling}{\darkblue{\RIGHTcircle}}

\definecolor{wheat1}{rgb}{1.000000,0.905882,0.729412}
\definecolor{LightGray}{rgb}{0.827451,0.827451,0.827451}

\newcolumntype{a}{>{\columncolor{wheat1}}l}

\definecolor{mygray}{rgb}{0.5,0.5,0.5}
\definecolor{mymauve}{rgb}{0.58,0,0.82}
\definecolor{darkblue}{rgb}{0.0,0.0,0.6}
\definecolor{darkgreen}{rgb}{0.0,0.6,0}
\definecolor{maroon}{RGB}{102, 0, 0}
\definecolor{Maroon}{cmyk}{0,0.87,0.68,0.32}
\definecolor{darkred}{RGB}{139, 0, 0}
\definecolor{forestgreen}{RGB}{34, 139, 34}

\newif\ifANNOYMIZE
\ANNOYMIZEfalse

\newif\ifACM
\ACMtrue  

\ifACM
\fi

\ifACM
\newcommand{\myfig}{Figure\xspace}
\else
\newcommand{\myfig}{Fig.\xspace}
\fi

\ifACM
\newcommand{\mysec}{\S}
\else
\newcommand{\mysec}{Section\xspace}
\fi

\newcommand{\names}{IABIs\xspace}
\newcommand{\name}{IABI\xspace}

\begin{document}

%
\title{On the Usability (In)Security of In-App Browsing Interfaces in Mobile Apps}

\ifANNOYMIZE
\author{Anonymous Submission}
\else
\author{Zicheng Zhang}
\affiliation{
  \institution{Singapore Management University}
  \country{Singapore}}
\email{zczhang.2020@phdcs.smu.edu.sg}

\author{Daoyuan Wu}
\authornote{Daoyuan Wu is the corresponding author.}
\affiliation{
  \institution{The Chinese University of Hong Kong}
  \country{Hong Kong, China}
}
\email{dywu@ie.cuhk.edu.hk}

\author{Lixiang Li}
\authornote{Lixiang Li was a MSc student when he conducted this study through the Advanced Research and Development Project course at The Chinese University of Hong Kong.}
\affiliation{
  \institution{miHoYo Co., Ltd.}
  \country{China}}
\email{lixiang.li@mihoyo.com}

\author{Debin Gao}
\affiliation{
  \institution{Singapore Management University}
  \country{Singapore}
  }
\email{dbgao@smu.edu.sg}
\fi


\begin{abstract}

Due to the frequent encountering of web URLs in various application scenarios (e.g., chatting and email reading), many mobile apps \red{build} their in-app browsing interfaces (\names) to provide a seamless user experience.
Although this achieves user-friendliness by avoiding the constant switching between the subject app and the system built-in browser apps, we find that \names, if not well designed or customized, could result in usability security \red{risks}.

In this paper, we conduct the first empirical study on the usability (in)security of in-app browsing interfaces in both Android and iOS apps.
Specifically, we collect a dataset of 25 high-profile mobile apps from five common application categories that contain \names, including \textit{Facebook} and \textit{Gmail}, and perform a systematic analysis \red{(not end-user study though)} that comprises eight carefully designed security tests and covers the entire course of opening, displaying, and navigating an in-app web page. 
During this process, we obtain three major security findings:
(1) about 30\% of the tested apps fail to provide enough URL information for users to make informed decisions on opening an URL;
(2) nearly all custom \names have various problems in providing sufficient indicators to faithfully display an in-app page to users, whereas ten \names that are based on Chrome Custom Tabs and SFSafariViewController are generally secure;
and (3) only a few \names give \red{warnings} to remind users \red{of the risk of inputting passwords} during navigating a \red{(potentially phishing) login} page. 

Most developers had acknowledged our findings but their willingness and readiness to fix usability issues are rather low compared to fixing technical vulnerabilities, which is a puzzle in usability security research.
Nevertheless, to help mitigate risky \names and guide future designs, we propose a set of secure \name design principles.

\end{abstract}

%
%



\maketitle

\section{Introduction}
\label{sec:intro}

Nowadays, mobile applications (or apps) are heavily used in our daily life.
Although most app functionalities are self-contained, it is not uncommon for users to open (external) web URLs in their app UIs (user interfaces).
  For example, a user may need to open a URL sent from her friends in a chat app like \textit{Whatsapp} or \textit{WeChat}, or need to open a URL embedded in an email when using \textit{Gmail}.
To satisfy such URL opening requirements in non-browser apps, one could offload the task to the system built-in browser apps.
While this is simple for developers, it hurts user-friendliness due to the potential constant switching from the subject app to a system browser app.
As a result, many high-profile apps choose to provide their own \textit{in-app browsing interfaces}, or \names, for a seamless user experience.

However, if not designed or customized well, \names could introduce serious usability security issues.
The major reason is that \names are typically simplified implementations of browsing interfaces lacking security indicators as opposed to ``full-service'' and standalone browsers with well-thought usability security designs.
For example, an \name may not display the full URL domain name or simply miss the HTTP(S) indicator.
Motivated by this intuition, we conduct the first empirical study in this paper on the usability (in)security of in-app browsing interfaces in both Android and iOS apps.
To this end, we collect and analyze a dataset of 25 high-profile mobile apps that contain \names, such as \textit{WeChat, Twitter, Gmail, LinkedIn}, and \textit{Reddit}.
To make our results representative, these apps are selected from five common app categories, including Chat, Social, Mail, Business, and News.

Atop this dataset, we perform a systematic analysis that comprises eight carefully designed security tests (T1$\sim$T8 in three categories) and covers the entire course of interacting with an in-app web page in \names including page opening, displaying, and navigating.
First, before a user opens a URL, we test whether the subject app provides sufficient URL information to enable end users to make informed decisions on opening the URL in a trustworthy manner (T1).
Second, after the web page is loaded, we test whether the \name provides enough security indicators for end users to validate the trustworthiness of the displayed page. 
This includes whether the URL itself (T2), an HTTPS (secure) indicator (T3), and an HTTP (insecure) warning (T4) are displayed in the title/address bar, whether a security alert is prompted for URLs with TLS errors (T5), and whether \names could defend against phishing URLs with a fake HTTPS lock icon (T6) and a long sub-domain name (T7).
Third, during navigation of the web page, we test whether \name could give a specific warning if the browsed page asks users to input passwords in a (potentially phishing) login form (T8).

Although our analysis focused on the apps' performance rather than a study with direct end users, our cross-platform analysis results show the following major security findings:
\begin{itemize}
\item About 30\% of the tested apps do NOT display the complete URL, thus fail to provide enough information for a user to trustfully open an URL.
  Most of these apps omit the scheme (HTTP or HTTPS), while two apps (\textit{Weibo} and \textit{Quora}) completely hide the URL. 
  Another 30\% of the apps, despite outputting the full URL, display additional favicon or title information, which enables attackers to craft a fake favicon/title to mislead users.
 
\item Nearly all custom \names have various problems in providing sufficient indicators to faithfully display an in-app page to users, whereas ten \names that are based on Chrome Custom Tabs or SFSafariViewController are generally secure.
  Specifically, among the 15 apps implementing their own \names, over half do not display the domain name in the address bar, and nearly none provide HTTP(S) indicators, which makes them fail to defeat phishing with a lock emoji or a long subdomain.

\item Only a few \names, from the \textit{QQ}, and \textit{QQ Mail} apps, give specific warnings to remind users of dangerous operations (e.g., password inputting) during navigating a login page.
\end{itemize}

To understand developers' reaction to our findings and to potentially provide our recommendations on fixing severe \name issues, we issued security reports to all affected apps (details in Section~\ref{sec:response}).
Most developers acknowledged our findings and agreed with our assessment.
In particular, \textit{Instagram} had fixed its issues as we reported 
and \textit{LinkedIn} would patch it in its future versions. 
However, we also found that developers' willingness and readiness to fix usability security issues are rather low compared to fixing technical vulnerabilities.
Specifically, they refused to recognize them as {\em vulnerabilities} and were not willing to patch or improve their risky \names.
Nevertheless, to help mitigate risky \names and guide future designs, we propose a set of secure \name design principles in Section~\ref{sec:goodDesign}.

To sum up, we make the following contributions in this paper:
\begin{itemize}
  \item \textit{(Problem and analysis in \S\ref{sec:backg}-\S\ref{sec:method})}
    We summarized the attack surfaces on interacting with an \name and performed a systematic analysis with eight security tests.

  \item \textit{(Measurement results in \S\ref{sec:evaluate})}
    We obtained cross-platform analysis results and their three major security findings by extensively testing 25 high-profile mobile apps.

  \item \textit{(Reporting and defense in \S\ref{sec:response}-\S\ref{sec:goodDesign})}
    We reported our findings to affected vendors and analyzed their responses. We further proposed a set of \name design principles.
\end{itemize}

\section{Background}
\label{sec:backg}

When we are using mobile apps to, e.g., chat with friends, read posts
on social platforms, or read emails, we often need to open a URL link.
In order to provide a ``one-stop'' service to keep users within the
app interface without the need of switching to a web browser, many
apps have implemented their own in-app browsing interfaces (\names)
which typically use the underlying browsing engines to load the web
content.

\begin{figure}[t!]
    
    \includegraphics[width=0.49\textwidth]{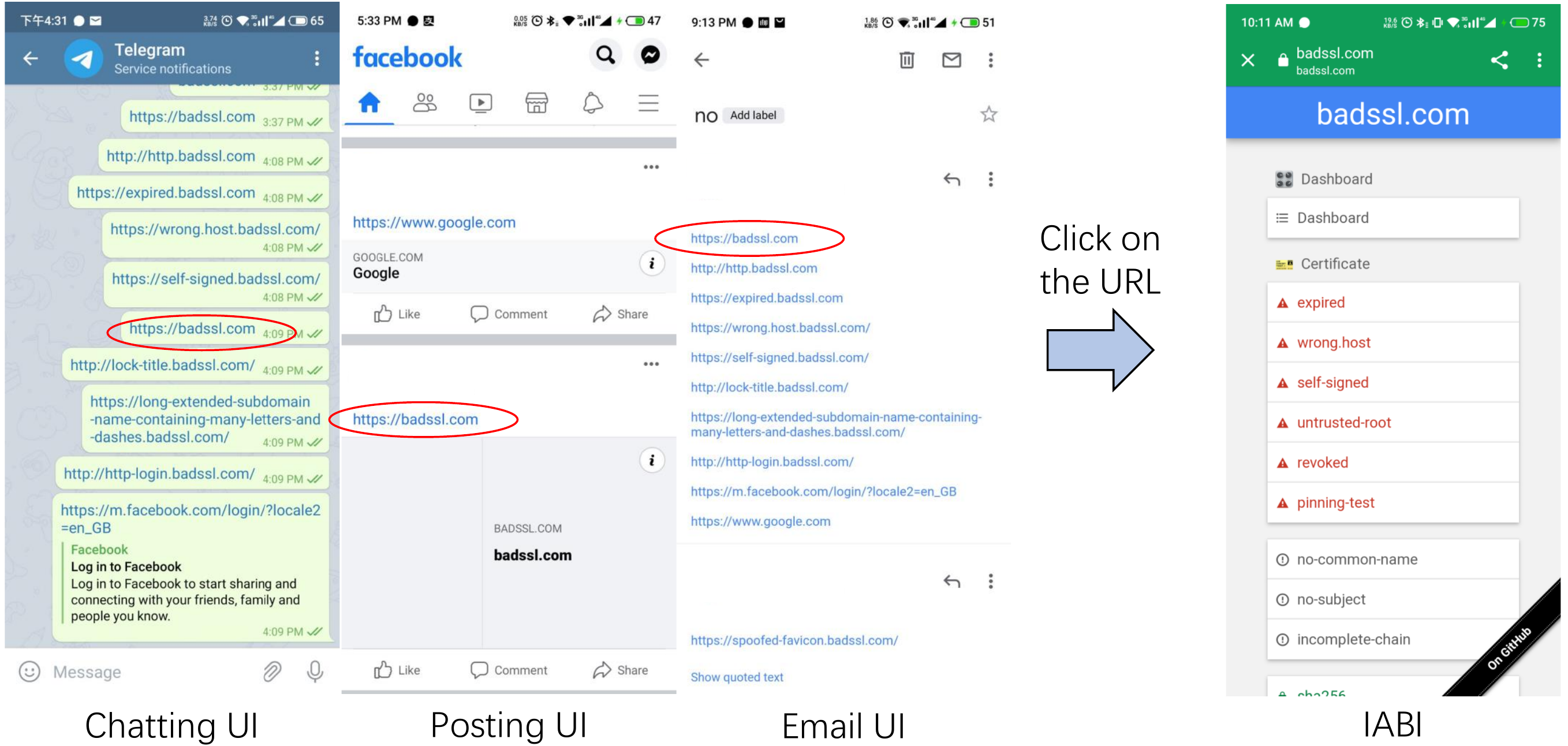}
    
    \vspace{-2ex}
    \caption{The process of opening a URL within an app.  It
      demonstrates the 3 sample situations in which we want to open a
      URL within an app, including chatting with friends (Chatting
      UI), posting on the social network (Posting UI), and reading or
      sending an email (Email UI).  When we click on the URL (e.g.,
      https://badssl.com), an app may use in-app browsing interfaces
      to open the URL.}
    \vspace{-4ex}
    \label{fig:background}
\end{figure}

Figure~\ref{fig:background} shows a typical process of 
opening a URL in an \name.  Sometimes the \name may also contain an address bar to display the information related to the web page.
Moreover, as shown in \myfig~\ref{fig:FlowChart}, apps have three ways to handle an URL request when a user clicks the URL.
First, some apps choose to jump out of the app and open a browser app to display the web page.
This situation is out of the scope of our paper because they do not contain in-app browsing interfaces.
The second way is to customize a browsing interface based on the Chrome Custom Tabs (CCT)~\cite{CCT} and SFSafariViewController (SF)~\cite{SFSafariViewController} libraries.
The third way is to create their own \names, which display the web page in the form of WebView (for
Android)~\cite{WebView} or UIWebView (for iOS)~\cite{UIWebView} instances.
Next, we explain the relevant terms in more details.


\textbf{IABI (in-app browsing interface) and its
address bar.}  We refer \red{to} the UI design of the entire screen
when a mobile app opens a web page as the In-App Browsing Interface
(\name).  In this paper, we mainly focus on the usability problems \red{of IABIs'} address bars.

\textbf{Underlying browsing engines.}  The implementation of
\names typically uses one of the following browsing engines:
\begin{itemize}
  \item Chrome Custom Tabs (CCT in Android)~\cite{CCT, ProgressiveWebApps18}
  \item SFSafariViewController (SF in iOS)~\cite{SFSafariViewController, ProgressiveWebApps18}
  \item WebView (in Android)~\cite{WebView, WebViewAtk11}
  \item UIWebView (in iOS)~\cite{UIWebView, MoST15} and
  \item Custom browser engines implemented in native code~\cite{FileCross14}.
\end{itemize}


\textbf{Chrome Custom Tabs and SFSafariViewController.}  CCT
is supported by Chrome, which is a web browser developed by Google. A
mobile app can send a special intent to Chrome to launch a CCT to open
websites without implementing a built-in browser engine by themselves,
assuming that Chrome is installed on the smartphone.  Chrome also
provides well-encapsulated APIs \red{for} developers \red{to make} some limited
browsing UI customization, \red{such as color and animation}.  Similarly, iOS has SF supported by Safari
for developers to incorporate into their apps easily.

\textbf{WebView and UIWebView.}  Although CCT and SF provide
convenient ways for a developer to implement IABI, one could choose,
instead, to use lower-level display engines WebView and UIWebView \red{(or even custom engines in native code~\cite{FileCross14})} for
more comprehensive control over the UI.  They allow developers to
monitor specific events (e.g., loading and navigating) upon triggering
of which developers can gather event information and make
corresponding responses.  The lower-level implementation of this
design provides developers with very flexible control of the UI, which
also implies more opportunities \red{for} design mistakes.







\section{Overview of Our Analysis}
\label{sec:method}

\begin{figure}[t!]
        \includegraphics[width=0.48\textwidth]{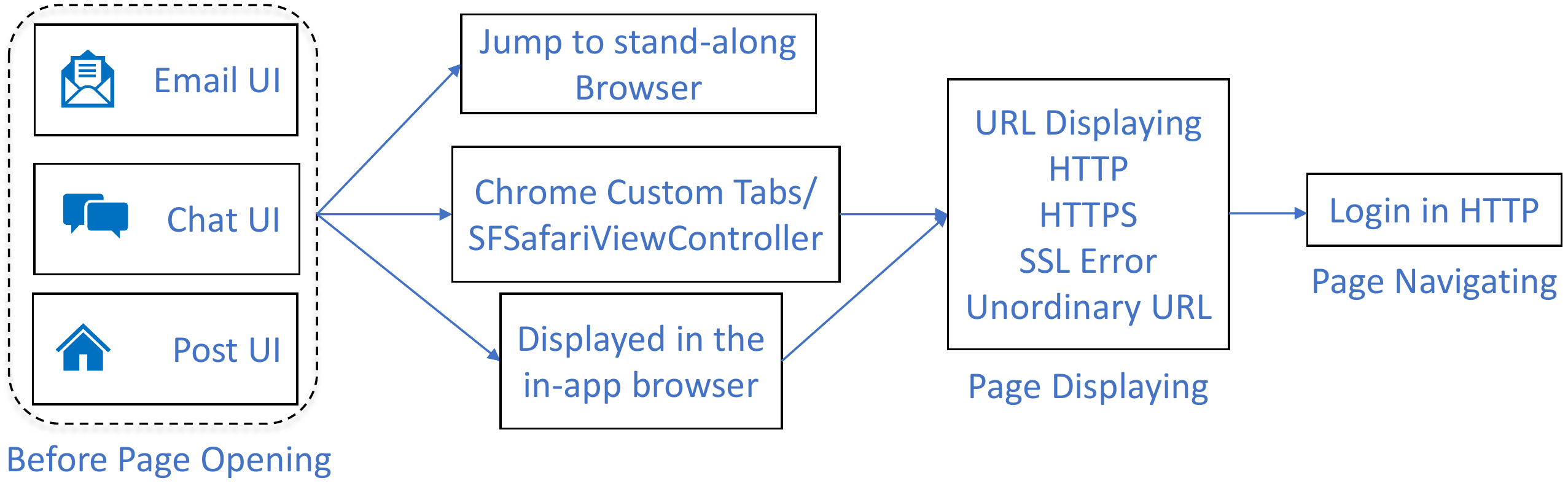}
    \vspace{-4ex}
    \caption{Three stages of interaction between an end user
      and the \name and their potential usability security risks,
      including (i) usability trust given before users \textit{open} a
      URL, (ii) security indicators to faithfully \textit{display} an
      in-app page to users, and (iii) specific warnings to
      remind users of dangerous operations during \textit{navigating} a login
      page.}
    \vspace{-4ex}
    \label{fig:FlowChart}
\end{figure}

To reveal the usability security issues of \names in real-world
applications, we analyze the \names in three phases corresponding to
opening, displaying, and navigating a web page; see
\myfig~\ref{fig:FlowChart}.  We design detailed security tests (T1
$\sim$ T8) to reveal security properties of the design of \names in
real-world applications.  URLs tested include those provided by
\url{https://badssl.com/} and homepages of Google and Facebook.

\begin{enumerate}
    \setlength{\topsep}{1ex}
    \setlength{\itemsep}{0pt}
          \setlength{\parsep}{0pt}
          \setlength{\parskip}{0pt}
    \item https://badssl.com
    \item http://http.badssl.com
    \item https://expired.badssl.com
    \item https://wrong.host.badssl.com
    \item https://self-signed.badssl.com
    \item http://lock-title.badssl.com
    \item https://long-extended-subdomain-name-containing-m
    any-letters-and-dashes.badssl.com
    \item http://http-login.badssl.com
    \item https://google.com
    \item https://m.facebook.com/login/
\end{enumerate}

In this section, we will explain the details of our security tests T1 $\sim$ T8 and then describe in detail how the individual tests are performed in each stage.

\subsection{Analyzing Risks before Page Opening}
\label{sec:pageOpen}

In the first phase of our analysis, we investigate how an app displays
URLs before users tap on them to open the web pages (\textbf{T1}).
The simplest design is to display only the URL of the website without
other content.  Some apps may show a box below the URL with
additional information about the web page, e.g., the title and favicon.

T1 is an indispensable step in the process of opening a web page in the app. Since it is not a part of the \name, we cannot compare it with mobile browsers, and existing work does not provide any principles about its design. 
Therefore, we assigned a GOOD rating according to criteria of other tests (T2, T6 \& T7, see Section~\ref{sec:pageDisplay}).
It could be counter-intuitive, but we find that only displaying
the URL without any other information could actually be a GOOD design
because any additional information displayed (e.g., favicon) could
potentially be taken advantage of by an attacker to provide misleading
information (e.g., favicon being a lock emoji).
NEUTRAL and BAD ratings are awarded accordingly.

To perform the test, we input all tested URLs to the subject apps and
check the corresponding display.  Note that at this moment the app
does not open the website yet. Some apps may pre-load the website to
get brief information about the website.  We discuss the results of T1 in Section~\ref{sec:urlRes}.



\subsection{Analyzing Risks on Page Displaying}
\label{sec:pageDisplay}

After an end user taps on the URL, the app could open the web page by
switching to a stand-alone web browser (out of scope of our paper) or
within the app implemented either with Chrome Custom Tabs
(CCT)/SFSafaraViewController (SF) or its customized \name.

When displaying the corresponding web content, different apps could
have their own design on the address bar, which is the
focus of the analysis in this stage.  It includes six tests as follows.
\begin{itemize}
  \item \textbf{T2}: whether the URL is shown on the address bar (using
    URL9). End users need this information to know the origin of the page. Various designs include showing the URL and/or domain
    name.
  \item \textbf{T3 and T4}: how HTTPS and HTTP protocols are handled in the URL (using URL1 and URL2). The HTTPS indicator is very intuitive for users to recognize whether a web page meets the TLS requirement or not. \names should display the corresponding indicator in both HTTP and HTTPS webpages. If an \name only displays the HTTPS indicators but shows no indicator for an HTTP page, the user may not know that the HTTP page is not secure before she opens another HTTPS page and sees the indicator.
  \item \textbf{T5}: how SSL errors are handled (using URL3$\sim$5).
    The SSL errors tested include expired certificates, wrong hosts,
    and self-signed certificates.  Various designs include blocking
    access, prompting options to end users, or accessing the web page
    without any warnings.
  \item \textbf{T6}: how the title with a lock emoji is displayed (using
    URL6).  Showing the lock emoji could mislead end users into believing that it is a secure web page with HTTPS protocol.
  \item \textbf{T7}: how URLs with long subdomain names are displayed
    (using URL7).  Displaying only the long subdomain without the
    domain name could present an illusion of visiting a trusted domain.
\end{itemize}

Our ratings for T2 $\sim$ T7 are based on the evaluation of security indicators and principles on mobile browsers in existing work~\cite{15TMC_indicator, TMC13SSL, ICS12measuring},
which perform systematic analysis based on best pratices outlined in the World Wide Web (W3C) guidelines~\cite{W3C}. In the following explanations, we first outline principles we extract from such guidelines that are applicable to \names, and then justify our definitions of the various ratings used in this paper.

\begin{enumerate}[leftmargin=2em, itemsep=0pt, label=\arabic*)]
    \item \textbf{Identity Signal: Availability.} The security indicators showing identity of a website MUST be available to the user through either the primary or secondary interface at all times. We believe that \names should at least display the domain name of a website which is the basic identity of a page. Therefore in \textbf{T2}, we assign a GOOD rating to displaying the URL or domain name of a web page.
    \item \textbf{Error messages: Interruption/Prceeding options/Inhibit interaction.} These three principles require that the error warnings MUST interrupt the users' current task and inhibit the user to interact with the destination website. Meanwhile, the warnings MUST provide the user with distinct options (MUST NOT be only to continue).
    Accordingly, in \textbf{T5} we test how \names react to erroneous certificates, like wrong-host, expired, or self-signed certificates. Our GOOD rating is consistent with this guideline, which is displaying a prompt with the option to continue or not before the user opens the SSL-error page. The only difference is that we relax the requirement a bit and allow \names to directly stop loading that page without providing options.
    A BAD rating is given to designs which directly open those pages with certificate errors. We assign NEUTRAL to designs that handover the issue to a standalone browser, which is ``lazy'' but not compromising security.
    \item \textbf{TLS indicator: Availability.} The TLS indicators MUST be available to the user through the primary or secondary interface at all times. Accordingly, we conduct \textbf{T3} and \textbf{T4}, which test whether the HTTPS and HTTP indicators are displayed on the address bar. A GOOD rating is to display the indicators (lock/exclamation icons) to show whether a page meets the TLS requirement. A BAD design does not show any indicators. Displaying the scheme (``https://'' or ``http://'') also serves as an indicator, although not as intuitive as icons and therefore receives a NEUTRAL score.
    \item \textbf{TLS indicator: Content and Indicator Proximity.} Content MUST NOT be displayed in a manner that confuses hosted content or browser indicators. In this paper, we conduct \textbf{T6} and \textbf{T7} to test whether the lock emoji in the titles and long sub-domain names could confuse the users. According to this guideline, the \names should not allow lock emoji to mimic the HTTPS indicators or allow the long sub-domain name to mislead users, which are the GOOD ratings of T6 and T7. We define NEUTRAL as displaying both the security indicators and the lock emoji (T6). In T7, NEUTRAL is not displaying the domain name (though it is a BAD design in T2), and a BAD design refers to those  ignoring this principle and allowing the two items to mislead users.
\end{enumerate}

The results of our analysis into T2 $\sim$ T7 are presented in
Section~\ref{RQ2:Page_Displaying}.

\subsection{Analyzing Risks on Page Navigating}
\label{sec:pageNavigate}

When a user navigates a web page in an app's \name, it is dangerous to input the
username and password information in a login form because \names are more vulnerable to phishing attacks than standalone browsers~\cite{CCS17hindsight}.
Therefore, well-designed \names should give \textit{specific} and \textit{extra} warnings to remind users of the risk of inputting passwords during navigating a login page.
Moreover, they should cover not only insecure HTTP login pages but also HTTPS login pages with vaild but illegitimate certificates. That is because an attacker can forge a phishing page with the title of a popular page (e.g., \texttt{alibaba.com}) by simply using a CA-issued (valid) certificate on a similar domain (e.g., \texttt{alibababa.com}) and thus meet the TLS requirement.

Based on this consideration, we test whether \names would show a \textit{specific} or \textit{extra} warning during navigation of a login page as compared with their normal behaviors on a non-login page (\textbf{T8}).
We conduct this test using URL8 and URL10, which are example HTTP and HTTPS login pages, respectively.
Note that for \textit{Facebook} and \textit{FB Messenger}, we use our university's HTTP login page since URL10 is already the Facebook login page.
For each \name, we navigate to the two login pages, input a username and a password, and check whether the \name shows a specific warning on our operation.

Specifically, we assign a GOOD rating if the subject \name provides specific warnings for both HTTP and HTTPS login pages. 
In contrast, we give the NEUTRAL and BAD ratings if the subject \name displays a warning for at least one login page or has no such warning for both pages, respectively.
We show the test results of T8 in Section~\ref{RQ3:Page_Navigating}.

\section{Cross-platform Analysis Results}
\label{sec:evaluate}

We begin this section with an overview of the test apps used in our
analysis (Section~\ref{sec:dataset}).  Section~\ref{sec:urlRes},
\ref{RQ2:Page_Displaying}, and \ref{RQ3:Page_Navigating} cover our
analysis results on \names handling of URLs before web page opening,
during web page opening, and during web page navigation, respectively.

\subsection{Test Apps and Overall Analysis Results}
\label{sec:dataset}

\begin{table}[t!]
\caption{Subject mobile apps we tested and \# of their Apple Store ratings and Google Play installs.}
\label{tab:Dataset}
\vspace{-2ex}
\scalebox{0.9}{
  \begin{tabular}{|c|c|c|c|}
    \hline
    \textbf{Category}         & \textbf{App Name} & \textbf{\# of Ratings} & \textbf{\# of Installs} \\ \hline
    \multirow{8}{*}{Chat}     & WeChat           & 4,000,000+             & 100M-500M               \\
                              & FB Messenger     & 1,000,000+             & 1B-5B                   \\
                              & QQ               & 700,000+               & 5B-10B                  \\
                              & Snapchat         & 300,000+               & 1B-5B                   \\
                              & LINE             & 200,000+               & 500M-1B                 \\
                              & Telegram         & 60,000+                & 100M-500M               \\
                              & KakaoTalk        & 60,000+                & 100M-500M               \\
                              & Hangouts         & 40,000+                & 1B-5B                   \\ \hline
    \multirow{6}{*}{Social}   & Instagram        & 10,000,000+            & 1B-5B                   \\
                              & Weibo            & 500,000+               & 1B-5B                   \\
                              & Facebook         & 400,000+               & 5B-10B                  \\
                              & Twitter          & 200,000+               & 500M-1B                 \\
                              & Tumblr           & 200,000+               & 100M-500M               \\
                              & VK Russia        & 200,000+               & 100M-500M               \\ \hline
    \multirow{4}{*}{Email}     & Gmail            & 100,000+               & 5B-10B                  \\
                              & 163 Mail         & 100,000+               & 50M-100M                \\
                              & Mail.ru          & 100,000+               & 50M-100M                \\
                              & QQ Mail          & 50,000+                & 100M-500M               \\ \hline
    \multirow{2}{*}{Business} & Alipay           & 600,000+               & 1B-5B                   \\
                              & LinkedIn         & 50,000+                & 500M-1B                 \\ \hline
    \multirow{5}{*}{News}     & Toutiao          & 2,000,000+             & 100M-500M               \\
                              & Reddit           & 1,000,000+             & 1M-50M                  \\
                              & Baidu            & 800,000+               & 1M-50M                  \\
                              & Zhihu            & 700,000+               & 100M-500M               \\
                              & Quora            & 70,000+                & 1M-50M                  \\ \hline
    \end{tabular}
}
\end{table}

Table~\ref{tab:Dataset} shows the category, number of ratings on Apple Store, and installs on Google Play of the 25 high-profile applications we use in
our analysis.  We reiterate that these apps are selected because they
have an \name and they are good representations of popular apps from
various usage categories.
In each category, we sort these apps according to their popularity on Apple Store because users in China typically do not use Google Play to install Android apps.

Table~\ref{tab:mainData} presents an overview of our analysis results,
with details discussed in Section~\ref{sec:urlRes},
\ref{RQ2:Page_Displaying}, and \ref{RQ3:Page_Navigating}.  Note that
when an app uses Chrome Custom Tabs (CCT) or SFSafariViewController
(SF) to implement \name, its behavior will always be the same (defined
in CCT and SF; see row ``CCT/SF'' in Table~\ref{tab:mainData}).
Therefore, our analysis and discussion below focus more on the
analysis of ``own \name'' implementation, where each subject app makes
its own call on the design and implementation.  As the results on
Android and iOS are the same in most subject apps, in the following
detailed discussions, we will first focus on our analysis on the
Android platform followed by a brief comparison with results on iOS.





\begin{table*}[t!]
  \centering
  \caption{Cross-platform \red{IABI} test results \red{in terms of their usability risks in opening, displaying, and navigating a web page}.}
\vspace{-2ex}
\label{tab:mainData}
\begin{center}
\scalebox{0.850}{
\arrayrulecolor{black}

\begin{tabular}{|c|c|c|c|c|c|c|c|c|c|c|c|}
  \hline
  \multirow{3}{*}{Category} & \multirow{3}{*}{App Name} & \multirow{3}{*}{CCT/SF} & \multirow{3}{*}{Own IABI} & \multirow{2}{*}{\begin{tabular}[c]{@{}c@{}}Before \\ opening\end{tabular}} & \multicolumn{7}{c|}{Detailed test results when the subject app uses its own \name}                                        \\ \cline{6-12} 
                            &                           &                         &                           &                                                                            & \multicolumn{6}{c|}{Displaying a web page}                                                              & Navigating     \\ \cline{5-12} 
                            &                           &                         &                           & T1: URL                                                                    & T2: URL        & T3: HTTPS      & T4: HTTP       & T5: SSL        & T6: Lock-title & T7: Sub-domain  & T8: Login      \\ \hline
  \multicolumn{2}{|c|}{CCT/SF}                          & * | *                   & * | *                     & * | *                                                                      & \gou | \gou    & \gou | \gou    & \gou | \gou    & \gou | \gou    & \ling | \gou   & \gou | \gou     & \cha | \ling    \\ \hline
  \multirow{8}{*}{Chats}    & WeChat                    & \cha | \cha             & \gou | \gou               & \gou | \gou                                                                & \cha | \cha    & \cha | \cha    & \cha | \cha    & \gou | \gou    & \cha | \cha    & \ling | \ling   & \cha | \cha    \\
                            & FB Messenger              & \cha | \cha             & \gou | \gou               & \ling | \ling                                                              & \gou | \gou    & \gou | \gou    & \cha | \cha    & \gou | \gou    & \cha | \gou    & \cha | \gou     & \cha | \cha    \\
                            & QQ                        & \cha | \cha             & \gou | \gou               & \gou | \gou                                                                & \cha | \cha    & \cha | \cha    & \cha | \cha    & \gou | \gou    & \cha | \cha    & \ling | \ling   & \gou | \gou    \\
                            & Snapchat                  & \cha | \cha             & \gou | \gou               & \cha | \cha                                                                & \gou | \gou    & \cha | \gou    & \cha | \cha    & \gou | \gou    & \cha | \cha    & \cha | \cha     & \cha | \cha    \\
                            & LINE                      & \cha | \cha             & \gou | \gou               & \ling | \ling                                                              & \gou | \gou    & \ling | \ling  & \ling | \ling  & \gou | \gou    & \ling | \ling  & \gou | \gou     & \cha | \cha  \\
                            & Telegram                  & \gou | \gou             & \cha | *                  & \gou | \gou                                                                & - | -          & - | -          & - | -          & - | -          & - | -          & - | -           & - | -          \\
                            & KakaoTalk                 & \cha | \cha             & \gou | \gou               & \ling | \ling                                                              & \gou | \gou    & \gou | \gou    & \cha | \cha    & \gou | \gou    & \gou | \gou    & \cha | \cha     & \cha | \cha    \\
                            & Hangouts                  & \gou | \gou             & \cha | *                  & \gou | \gou                                                                & - | -          & - | -          & - | -          & - | -          & - | -          & - | -           & - | -          \\ \hline
  \multirow{6}{*}{Social}   & Instagram                 & \cha | \cha             & \gou | \gou               & \ling | \ling                                                              & \gou | \gou    & \cha | \cha    & \cha | \cha    & \gou | \gou    & \cha | \cha    & \cha | \cha     & \cha | \cha    \\
                            & Weibo                     & \cha | \cha             & \gou | \gou               & \cha | \cha                                                                & \cha | \cha    & \cha | \cha    & \cha | \cha    & \gou | \gou    & \cha | \cha    & \ling | \ling   & \cha | \cha    \\
                            & Facebook                  & \cha | \cha             & \gou | \gou               & \ling | \ling                                                              & \gou | \gou    & \gou | \gou    & \gou | \gou    & \gou | \gou    & \gou | \gou    & \cha | \cha     & \cha | \cha    \\
                            & Twitter                   & \gou | \gou             & \gou | *                  & \cha | \cha                                                                & (o)\gou | -    & (o)\ling | -   & (o)\ling | -   & (o)\ling | -   & (o)\ling | -   & (o)\cha | -     & (o)\cha | -   \\
                            & Tumblr                    & \gou | \gou             & \cha | *                  & \cha | \cha                                                                & - | -          & - | -          & - | -          & - | -          & (t)\gou | -    & - | -           & - | -          \\
                            & VK Russia                 & \gou | \gou             & \cha | *                  & \gou | \ling                                                               & - | -          & - | -          & - | -          & - | -          & - | -          & - | -           & - | -          \\ \hline
  \multirow{4}{*}{Email}    & Gmail                     & \gou | \gou             & \cha | *                  & \gou | \gou                                                                & - | -          & - | -          & - | -          & - | -          & - | -          & - | -           & - | -          \\
                            & 163 Mail                  & \cha | \cha             & \gou | \gou               & \gou | \gou                                                                & \cha | \cha    & \cha | \cha    & \cha | \cha    & \gou | \gou    & \cha | \cha    & \ling | \ling   & \cha | \cha    \\
                            & QQ Mail                   & \cha | \cha             & \gou | \gou               & \gou | \gou                                                                & \cha | \cha    & \cha | \cha    & \cha | \cha    & \gou | \gou    & \cha | \cha    & \ling | \ling   & \gou | \gou    \\
                            & Mail.ru                   & \gou | \gou             & \cha | *                  & \gou | \gou                                                                & - | -          & - | -          & - | -          & - | -          & - | -          & - | -           & - | -          \\ \hline
  \multirow{2}{*}{Business} & Alipay                    & \cha | \cha             & \gou | \gou               & \gou | \gou                                                                & \cha | \cha    & \cha | \cha    & \cha | \cha    & \cha | \gou    & \cha | \cha    & \ling | \ling   & \cha | \cha    \\
                            & LinkedIn                  & \gou | \cha             & \gou | \gou               & \ling | \ling                                                              & (o)\cha | \gou & (o)\cha | \cha & (o)\cha | \cha & (o)\gou | \gou & (o)\cha | \cha & (o)\ling | \cha & (o)\cha | \cha \\ \hline
  \multirow{5}{*}{News}     & Toutiao                   & \cha | \cha             & \gou | \gou               & \cha | \cha                                                                & \cha | \cha    & \cha | \cha    & \cha | \cha    & \gou | \gou    & \ling | \ling  & \ling | \ling   & \cha | \cha    \\
                            & Reddit                    & \gou | \gou             & \gou | *                  & \gou | \gou                                                                & (o)\gou | -    & (o)\gou | -    & (o)\cha | -    & (o)\gou | -    & (o)\gou | -    & (o)\cha | -     & (o)\cha | -    \\
                            & Baidu                     & \cha | \cha             & \gou | \gou               & \gou | \gou                                                                & \cha | \cha    & \cha | \cha    & \cha | \cha    & \gou | \gou    & \ling | \ling  & \ling | \ling   & \cha | \cha    \\
                            & Zhihu                     & \cha | \cha             & \gou | \gou               & \cha | \cha                                                                & \cha | \cha    & \cha | \cha    & \cha | \cha    & \cha | \gou    & \cha | \cha    & \cha | \cha     & \cha | \cha    \\
                            & Quora                     & \gou | \gou             & \cha | *                  & \cha | \cha                                                                & - | -          & - | -          & - | -          & - | -          & (t)\gou | -    & - | -           & - | -          \\ \hline
  \end{tabular}
}
  \begin{center}
  {\raggedright \small Column ``CCT/SF'': whether the subject app uses Chrome Custom Tabs (Android) or SFSafariViewController (iOS) to implement \name.
  \\ Column ``Own \name'': whether the subject app uses its own implementation of \name.  Note that an app could present both CCT/SF and Own \name behavior (when the phone has and does not have Chrome/Safari installed, respectively).  Since Safari is installed by default on iOS, the app's ``own \name'' status is unknown when it already has SF implementation.
  \\ Row ``CCT/SF'': analysis results when using CCT/SF to implement \name.  Note that all subject apps will present the same behavior when using CCT/SF.
  \\ Cell format: $<$analysis result on Android$>$ | $<$analysis result on iOS$>$
  \\ Symbols: \gou = Good or Yes; \ling = Neutral; \cha = Bad or No; \red{`*' = Unknown; `-' =  Same as CCT/SF; `(o)' = Using \textbf{(o)}wn \name if no Chrome; `(t)' = CCT without \textbf{(t)}itle}.
  \par}
  \end{center}
\end{center}
\end{table*}

\subsection{Usability Risks before Page Opening}
\label{sec:urlRes}

As discussed in Section~\ref{sec:pageOpen}, this part of the
analysis concerns how the URL is displayed before end users opens it.

\subsubsection{T1: Displayed URLs before page opening}
\label{RQ1:DisplaysingURLs}

\begin{figure}[t!]
    \includegraphics[width=0.4\textwidth]{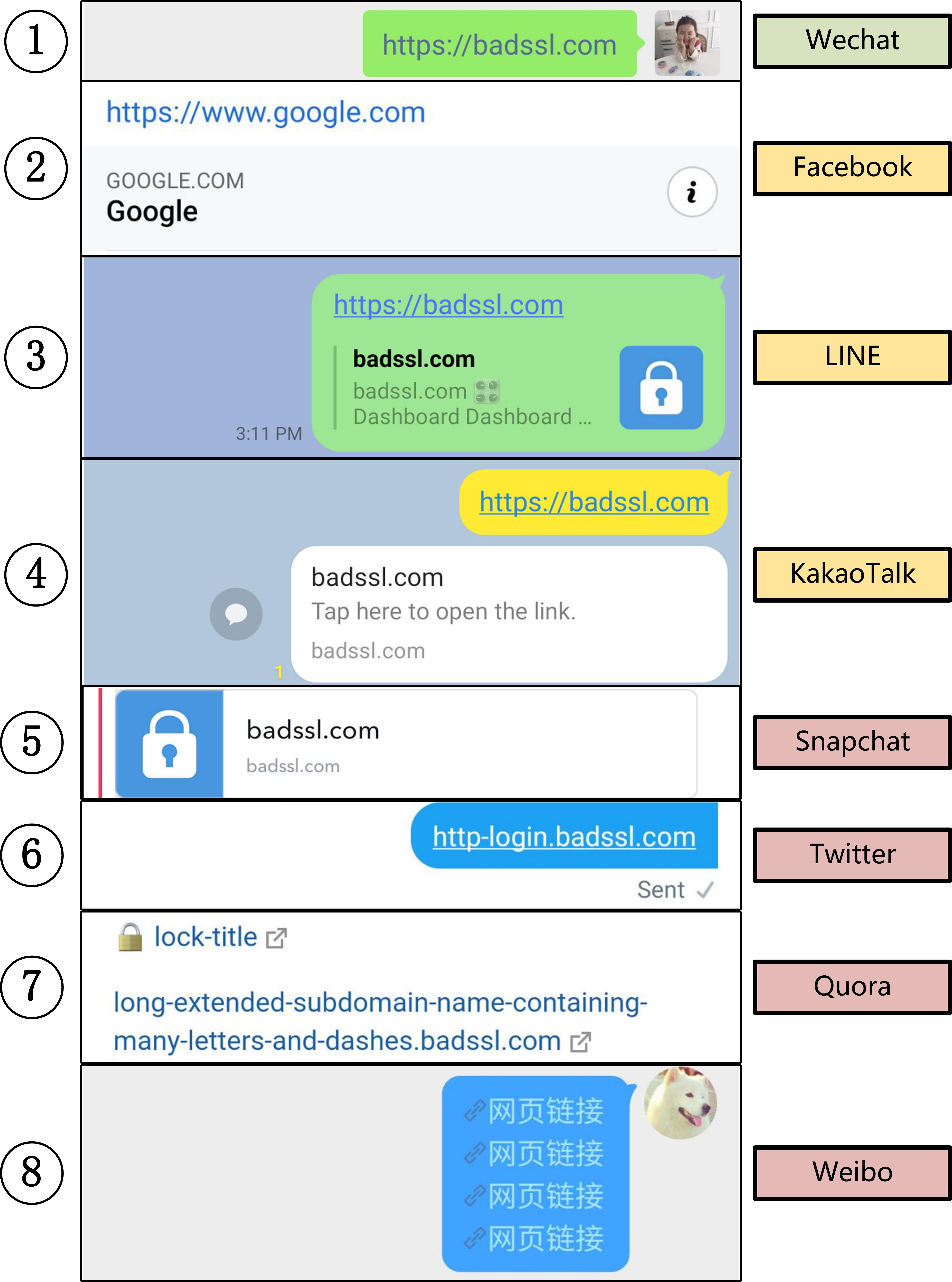}
    \vspace{-2ex}
    \caption{Examples of displayed URLs before page opening (T1).}
    \vspace{-4ex}
    \label{fig:URLDisplay}
\end{figure}

\myfig~\ref{fig:URLDisplay} provides screenshots of a few
representative apps on how they display a URL before end users click
on it.  After analyzing the different handling, we categorize them
into three buckets --- Good, Neutral, and Bad\footnote{We notice cases
  in which a subject app displays URLs in different ways in different
  activities (e.g., in a chat window and in a wall/post window), in
  which our analysis result is based on its worst-case
  categorization. }.


\textbf{GOOD.} The most common way (accounts for roughly 50\% of our
subject apps) is \red{displaying} the complete URL; see Case 1 in
\myfig~\ref{fig:URLDisplay}.  We consider this a \red{GOOD} practice as end
users can see the full URL without being misled by maliciously crafted
favicons or titles (see the BAD cases later).

\textbf{NEUTRAL.} Some apps may display additional information of the
corresponding web page (see Case 2, 3, and 4 in
\myfig~\ref{fig:URLDisplay}; we confirm that they show the complete URL even if the URLs contain sub-domains and additional
  paths) including title, domain, favicon, and
even some page content.
Although such additional information might help in a legitimate URL, \red{it} could be also
misleading in other cases.
For example, a fake lock favicon is displayed in the case of \textit{LINE} in
\myfig~\ref{fig:URLDisplay}.  That said, apps in this category also
display the complete URL for inspection by end users; we therefore
consider the practice of displaying the complete URL together with
additional (potentially misleading) information as NEUTRAL.

\textbf{BAD.} This category refers to apps that do not display the
complete URL (including cases with missing HTTP/HTTPS scheme).  For example,
\textit{Snapchat} (Case 5) only displays the title, favicon, and
domain name (and see the lock-looking favicon).  It presents an
example in which end users are given no information about the actual
security of the URL.
\textit{Twitter} and \textit{Quora} (Case 6 and
7) have the scheme of the URL stripped off with only the domain name
left.  \textit{Weibo} (Case 8) displays an identical label for every
URL, which is also \red{BAD} in failing to provide adequate information of
the URL.

\textbf{iOS.} Most of the apps perform exactly the same on Android and
iOS in this analysis, with the exception of \textit{VK Russia} whose
Android version displays only the complete URL (GOOD) while its iOS
version displays the title and domain name as well (NEUTRAL).



\begin{center}
  \shadowbox{\parbox{0.45\textwidth}{
  \textbf{\small Takeaway in \S\ref{sec:urlRes}: About 30\% of the subject apps do NOT display the complete URL, failing to provide the necessary security indicators.  Most of them \red{omit} the scheme (HTTP
  or HTTPS), while \red{two apps completely hide} the URL content.
  Another 30\% of the apps, despite outputting the full URL, display
  additional favicon or title information, which potentially enables
  attackers to maliciously craft a fake favicon/title to mislead end
  users.\normalsize}
  }
}
\end{center}
		





\subsection{Usability Risks on Page Displaying}
\label{RQ2:Page_Displaying}

As discussed in Section~\ref{sec:pageDisplay}, this part of the
analysis focuses on the title and address bar developers typically add
to enhance the user interface of the app.  We first examine the
display of URL when loading the web page (T2), and then test 4 types
of URLs with HTTP (T3), HTTPS (T4), SSL errors (T5), and special
cases (T6 and T7).

\subsubsection{T2: Displayed URLs during page opening}
\label{RQ2:DisplayURLs}

\begin{figure}[t!]
    \includegraphics[width=0.4\textwidth]{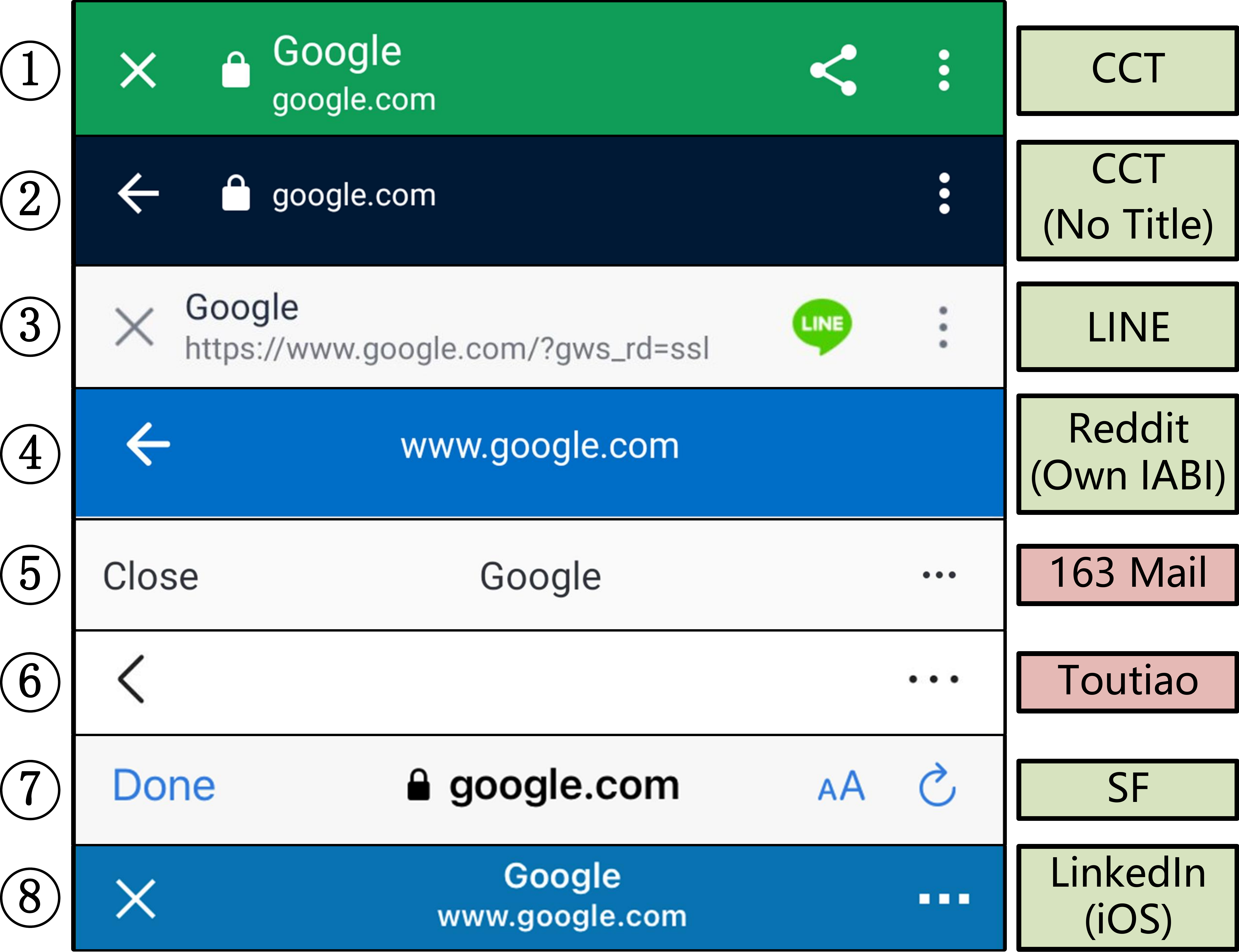}
    \caption{Examples of displayed URLs when a page is loaded (T2).}
    \vspace{-4ex}
    \label{fig:URLTitle}
\end{figure}

The importance of proper display of URLs here is similar to that in
T1, with three notable differences.  First, T2 focuses on the display
of URL only and leaves the analysis of scheme indicators to
T3$\sim$T5.  Second, web page redirection places an additional demand
on the display of URL\red{s} while pages are being opened.  Third, a preview
of the page no longer adds any usability or functionality since the
actual page is now opened.  Results on representative apps are shown
in Figure~\ref{fig:URLTitle}.

\textbf{Chrome Custom Tabs and SFSafariViewController.}  We first take
a look at how full-fledged browsers handle URL displays.  Both Chrome
Custom Tabs (CCT) and SFSafariViewController (SF) always display the
domain name in their address bars, which is considered a GOOD design.
Customization of CCT allows showing or hiding the title (using API
\texttt{setShowTitle(true)}); see Case 1 and 2 in
Figure~\ref{fig:URLTitle}).  10 of the 25 subject apps use CCT while 9
of them use SF.

We now turn our attention to \name with each application's own design
and implementation without using CCT/SF.  Note that not all subject
apps choose to provide their own \name implementation; see column
``Own \name'' in Table~\ref{tab:mainData}.  Similar to what we did
with T1, we categorize these implementations into buckets of GOOD,
NEUTRAL, and BAD, and present our findings on the Android platform
first followed by a comparison with corresponding iOS apps.

\begin{figure}[t!]
      \includegraphics[width=0.4\textwidth]{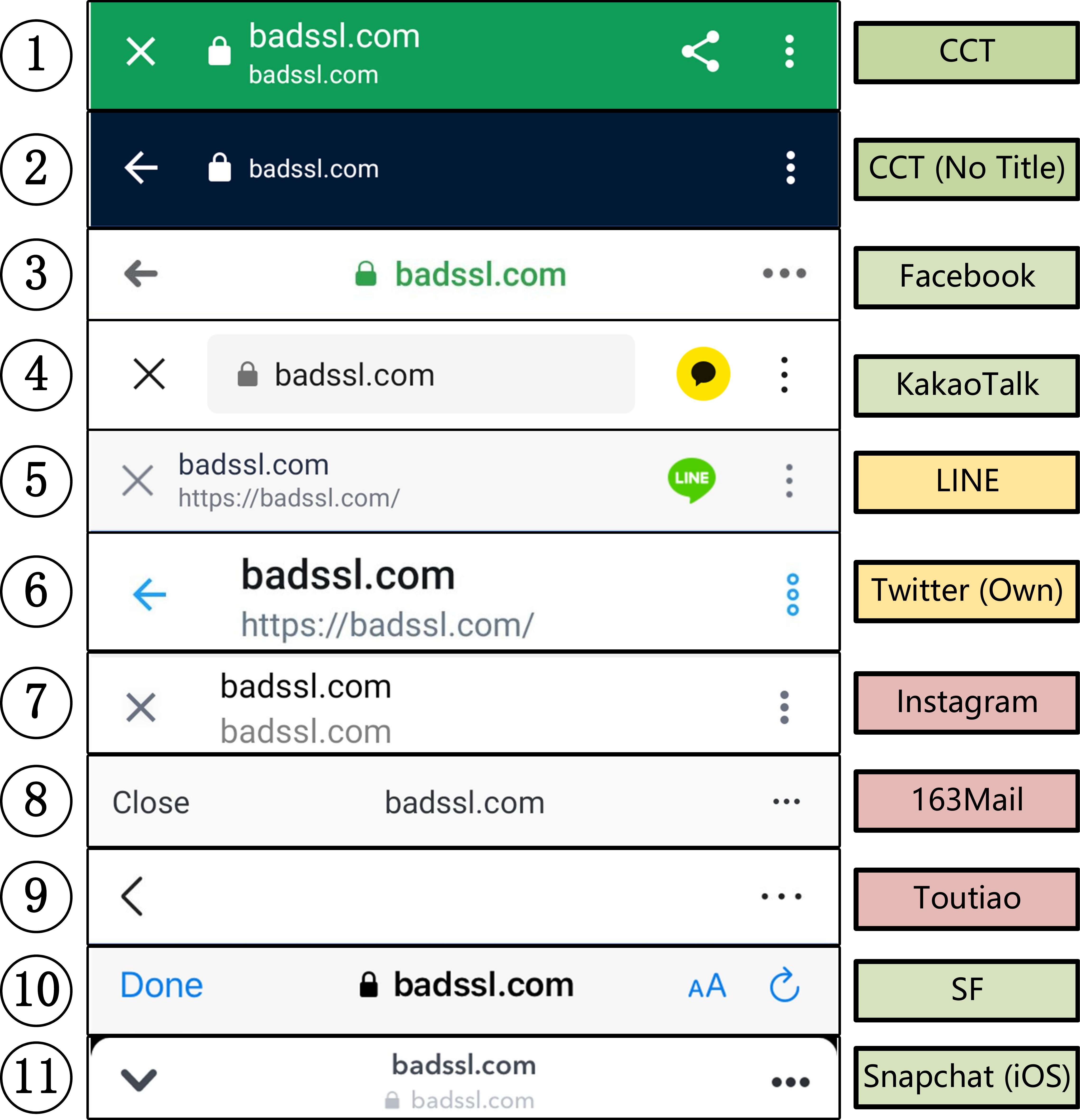}
      \vspace{-2ex}
      \caption{Examples of HTTPS indicators (T3). \red{`(Own)' refers to the app's own \name, similarly hereinafter}.}
      \vspace{-3ex}
      \label{fig:HTTPS}
  \end{figure}

\textbf{GOOD.}  We consider it a \red{GOOD} design provided that either 
complete URL or its domain is displayed.  A total of 8 Android apps
satisfy this requirement (e.g., Case 3 in Figure~\ref{fig:URLTitle})
out of 18 subject apps\footnote{Three apps (\textit{Twitter, LinkedIn} and \textit{Reddit}) provide both CCT/SF and own
  \name implementations.} that provide their own \name
implementations.

\textbf{BAD.} The other 10 subject apps only display the title of the
web page without the URL (or domain name), e.g., \textit{163Mail}
(Case 5), or even no title/address bar at all, e.g., \textit{Baidu},
which is a \red{BAD} design for the same reasons discussed in T1.  An
interesting observation is that 9 out of these 10 subject apps are from
China.

\textbf{Page redirection.} Web page redirection is common, and \names
shall always display information of the URL of the final landing page
being opened.  Our evaluation shows that all the subject apps pass
this test, should they use \texttt{WebView.getUrl()} to directly
retrieve the URL or use the arguments of a set of hook functions
within \texttt{WebViewClient} (\texttt{onPageStarted()} and
\texttt{onPageFinished()}).

\textbf{iOS.} On iOS, LinkedIn displays the title and domain name on
the address bar (Case 8), which is GOOD (as opposed to its \red{BAD} design
on the Android counterpart).  Other apps have exactly the same
performance on the two mobile platforms.


\subsubsection{T3: HTTPS Indicators}
\label{RQ2:HTTPSURLS}

Apps typically provide HTTPS indicators in the form of text (``https''
in the URL) or a lock icon.  Screenshots of representative apps in
this analysis are shown in \myfig~\ref{fig:HTTPS}.

\textbf{Chrome Custom Tabs and SFSafariViewController.} Both CCT and
SF use a lock icon as \red{the indicator} of HTTPS, which is not customizable
or removable by the app developers (Case 1 and 2 in
Figure~\ref{fig:HTTPS}).  We consider them \red{GOOD} designs.


\textbf{GOOD.} Similar designs can be found in own \name
implementations in three apps, \textit{Facebook} (Case 3), \textit{FB Messenger} and \textit{KakaoTalk} (Case 4).

\textbf{NEUTRAL.} Some apps rely on the scheme portion of the full
URL (the text ``https'') to indicate that HTTPS protocol is being used
(Case 5 and 6), which might not be very intuitive but suffice for
advanced users.

\textbf{BAD.} \red{The absence} of any HTTPS indicators (text or lock icon) is a
\red{BAD} design.  Surprisingly, we found 12 out of the 18 subject Android
apps with \red{their} own \name implementation falling into this category,
including \textit{Instagram} developed by the Facebook company
(\textit{Facebook} and \textit{FB Messenger} are GOOD though).


\textbf{iOS.} The iOS version of \textit{Snapchat} displays a lock
icon in its own IABI implementation, while its Android version does
not have any indicators.  Other apps have identical behavior with
respect to T2 on Android and iOS.

\subsubsection{T4: HTTP Indicators}
\label{RQ2:HTTPURLs}

\begin{figure}[t!]
    \includegraphics[width=0.4\textwidth]{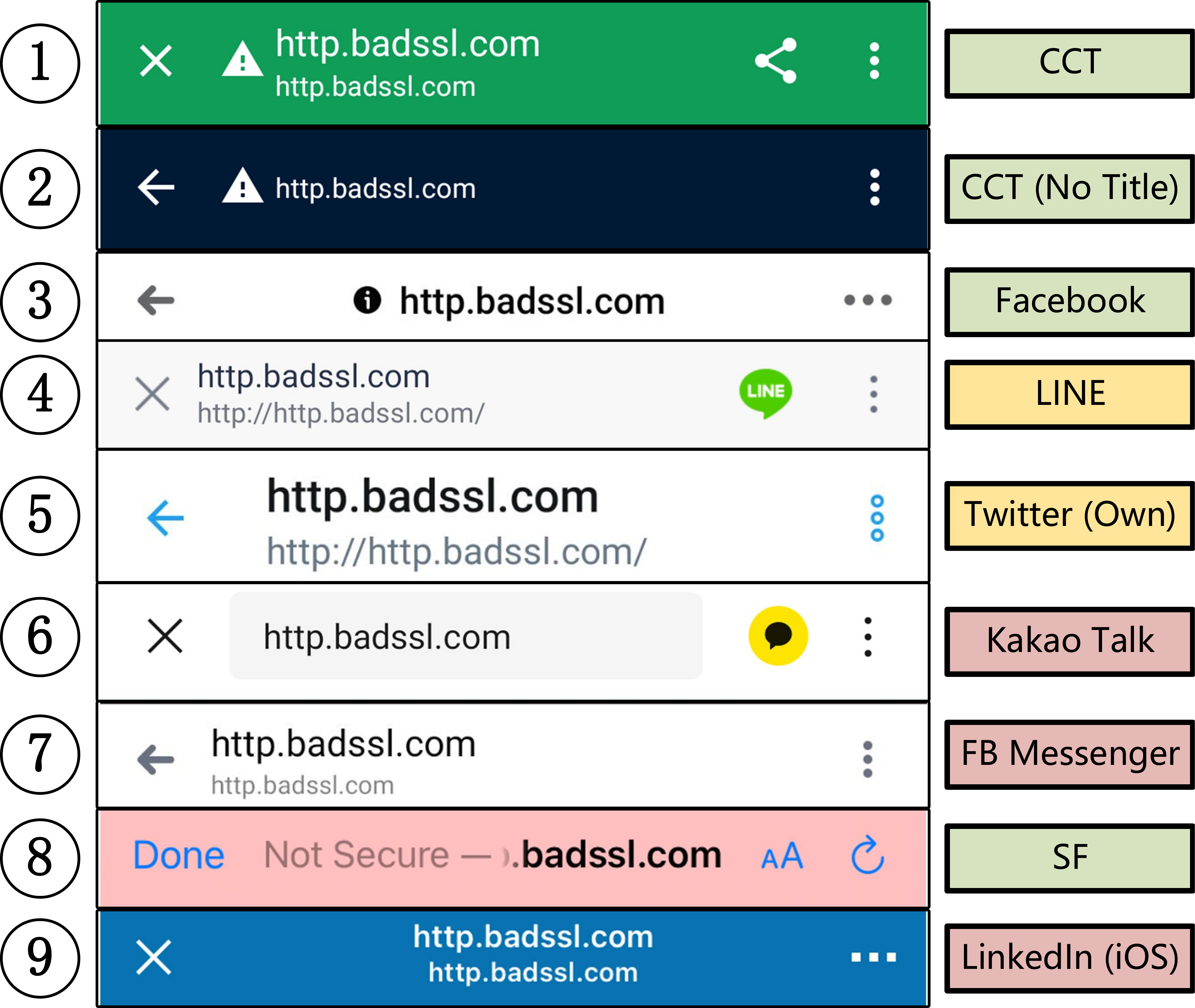}
    \vspace{-2ex}
    \caption{Examples of HTTP indicators (T4).}
    \vspace{-4ex}
    \label{fig:HTTP}
\end{figure}

As discussed in Section~\ref{sec:pageDisplay}, proper indicators for
HTTPS should not exempt an app from displaying an HTTP
indicator/prompt.  In other words, an HTTP indicator should always be
displayed regardless of \red{the} presence or absence of HTTPS indicators.
\myfig~\ref{fig:HTTP} shows screenshots of representative apps in this
analysis.

\textbf{Chrome Custom Tabs and SFSafariViewController.} CCT uses an
exclamation mark icon in place of the lock icon when the URL does not
meet TLS requirements; see Case 1 and 2 in \myfig~\ref{fig:HTTP}.
This icon is very intuitive and can give the user a clear warning.
Same as the lock icon for HTTPS, this exclamation mark icon cannot be
customized or removed by a developer.  SFSafariViewController uses the
text ``Not Secure'' as the indicator; see Case 8. \red{Both of them scored GOOD in this test.}

\textbf{GOOD.} Similar to the analysis of HTTPS indicators, a \red{GOOD}
design should always display an insecure indicator for HTTP.
Unfortunately, \textit{Facebook} is the only app scoring GOOD design
in this test (Case 3).

\textbf{NEUTRAL.} Displaying the complete URL with the scheme
portion of text ``http'' also serve\red{s} the purpose with a less intuitive
interface, and is considered a neutral design in our analysis.
\textit{LINE} and \textit{Twitter} join this category (Case 4 and 5).

\textbf{BAD.} \red{The absence} of any HTTP indicator is considered a \red{BAD}
design, and we have 15 out of 18 apps with their own \name
implementation in this category including \textit{FB Messenger}.  This
is an alarming finding.


\textbf{iOS.} Similarly, \textit{Facebook} is the only app with GOOD
design here with respect to HTTP indicators on iOS apps.

\subsubsection{T5: Certificate Errors}
\label{RQ2:CertificateError}

\begin{figure}[t!]
    \includegraphics[width=0.5\textwidth]{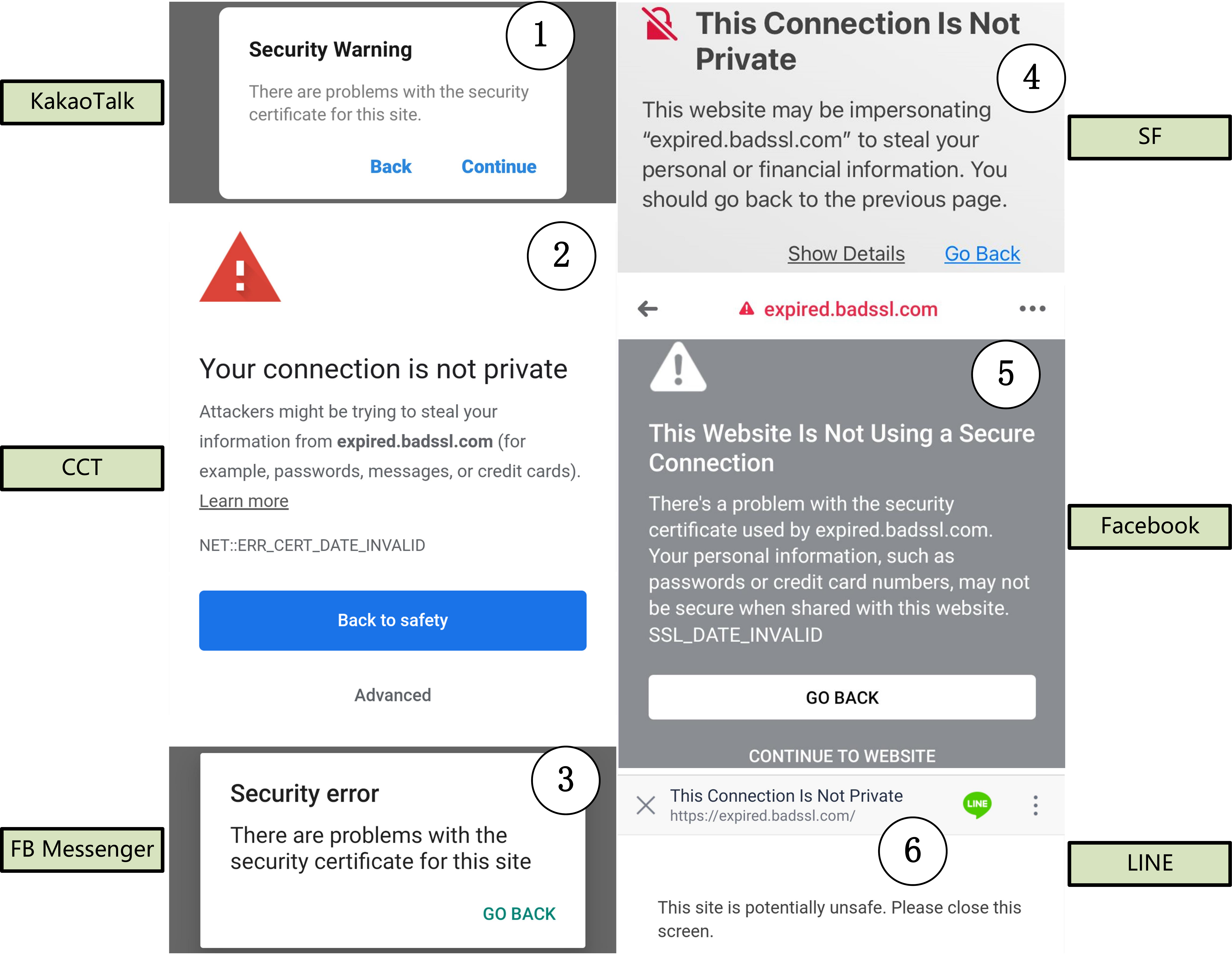}
    \vspace{-2ex}
    \caption{Examples of displaying expired certificates (T5).}
    \label{fig:Expired}
\end{figure}

As discussed in Section~\ref{sec:pageDisplay}, an app is supposed to
inform end users upon certificate errors (with expired or self-signed
certificates, or with certificates of the wrong host).  We test all
subject apps with URLs that contain such certificate errors, and
examine their corresponding prompts; see
\myfig~\ref{fig:Expired} with expired certificates as examples.
 
\textbf{Chrome Custom Tabs and SFSafariViewController.} Since most
layman end users do not possess the necessary technical background to make
informed decisions when they are prompted with a certificate error,
both CCT and SF (and the corresponding full-fledged browsers) introduce
``twisted'' routes for end users to proceed opening web pages with
certificate errors.  Cases~2 and~4 in \myfig~\ref{fig:Expired} show
examples of the design of such ``twisted'' routes, where an end user
will have to choose ``Advanced'' or ``Show Details'' before they are
given the option to continue browsing.  We consider these \red{GOOD} designs
with best security usability practice.

That said, both CCT and SF choose to remember such end user decisions
{\em across all apps with CCT/SF implementation}.  In other words, an
app with CCT/SF implementation would skip the certificate error
warning if a user had chosen to proceed with browsing the same
URL on any other CCT/SF apps. Although the CCT/SF implementation is
generally considered \red{GOOD}, such a design choice sacrifices security
for usability.

\textbf{GOOD.} Here we relax the security requirement and consider
apps that either refuse to open the page or prompt end users with
various options as GOOD designs.

With such relaxation on the definition of GOOD designs, all but 3
subject apps with \red{their} own \name implementations are in this category.
Among them are 7 apps that simply refuse to open the page, and 8 apps
that prompt the end users and give options to proceed.  There are two
interesting observations worth noting when we dig deeper into the
latter group.

First, 3 out of the 8 apps show details of specific certificate errors
(e.g., \textit{Facebook}, Case~5) while the other 5 apps skip the
details of errors (e.g., \textit{KakaoTalk}, Case~1).  Second, all
these apps remember the user selection and would not display any SSL
error indicators after proceeding to open the web page, except
\textit{Facebook} (case 5) which turns the domain name into red color and
include a red exclamation mark icon even after end user chooses to
proceed.

\textbf{NEUTRAL.} \textit{Twitter} instead chooses to launch the
system default browser to handle the web pages with certificate
errors.  We consider this design acceptable (NEUTRAL) although the
burden is now shifted to the default browser.

\textbf{BAD.} Ignoring the certificate errors or directly opening the
insecure web page are both considered as BAD designs.  In this category
we have \textit{Alipay} which directly opens a web page with wrong
host certificates and \textit{Zhihu} which indiscriminately displays
a prompt of visiting external websites regardless of presence of
absence of certificate errors.

\textbf{iOS.} Surprisingly, all iOS subject apps deliver \red{GOOD} designs
including \textit{Alipay} and \textit{Zhihu} whose Android versions
are BAD.  Both two iOS apps show blank
pages when attempting to open pages with certificate errors.  We
suspect that this better behavior of iOS apps is due to stricter
control for certificate errors on the iOS platform.



\begin{figure}[t!]
  \includegraphics[width=0.4\textwidth]{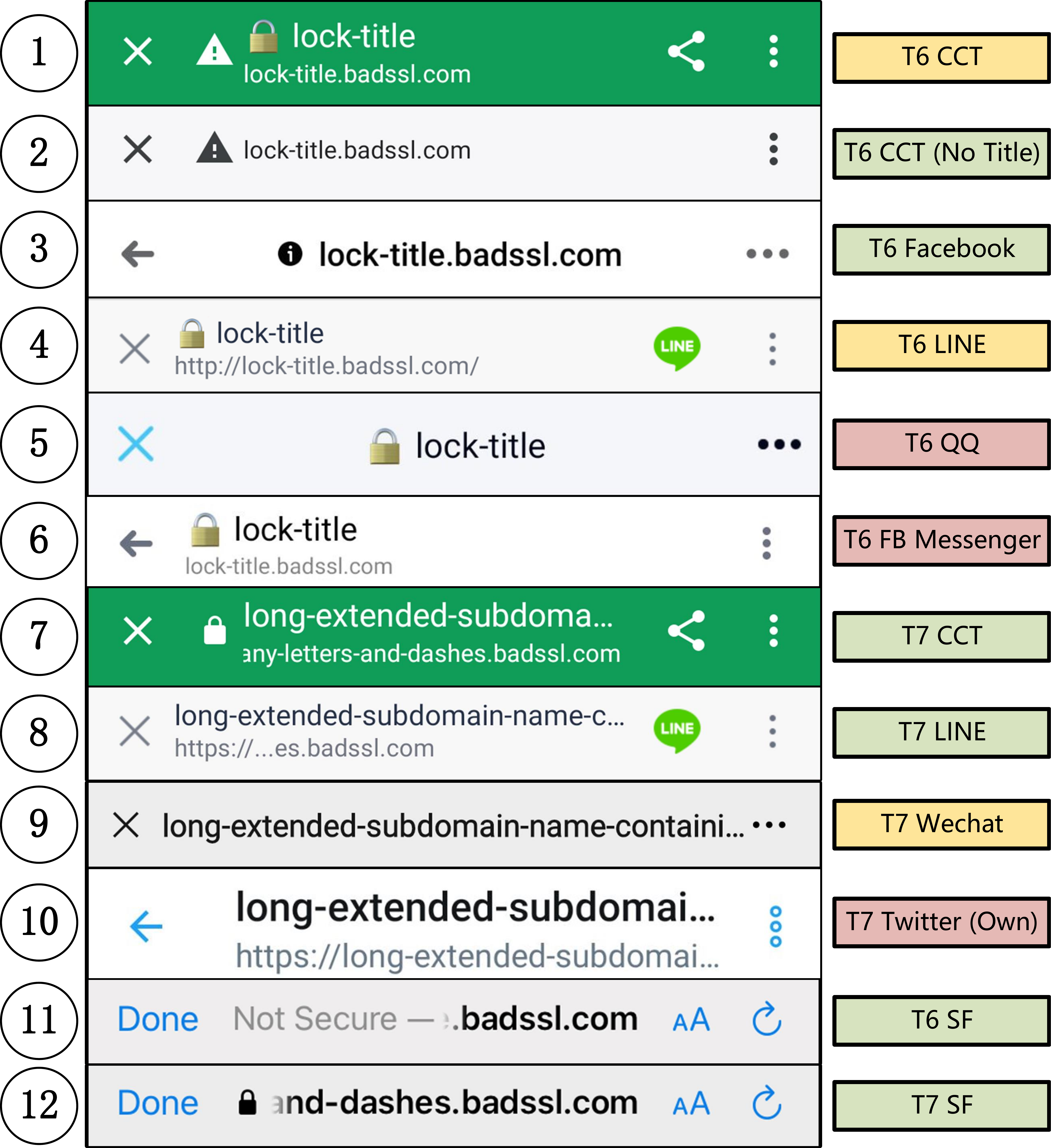}
  \vspace{-2ex}
  \caption{Examples of displaying special URLs (T6 \& T7).}
  \label{fig:UI}
\end{figure}

\subsubsection{T6 \& T7: Special URLs}
\label{RQ2:Unordinary URLs}

As discussed in Section~\ref{sec:pageDisplay}, T6 and T7 concern
about special URLs where a lock emoji is part of the title, and
where extended sub-domain names are used, respectively.
\myfig~\ref{fig:UI} shows screenshots of representative apps when they
process such special URLs.

\textbf{Chrome Custom Tabs and SFSafariViewController.}  CCT could be
configured with or without \red{the} title of the page being visible.  When the 
title is visible (Case 1 of \myfig~\ref{fig:UI}), the lock emoji (part
of the title) and the exclamation mark icon (due to the HTTP protocol
being used) show up next to \red{each other}, which is \red{confusing} but not
too bad as the warning is there.  In absence of the page title, either
due to developer configuration in CCT (Case 2) or due to SF in use
(Case 11), the HTTP warning is there without any confusion.

Regarding T7, both CCT and SF display the {\em suffix} of the URL
(with the long subdomain trimmed), which means that both CCT and SF
are immune to T7 attacks (see Cases~7 and~12).

\textbf{GOOD.} Regarding T6, a potential \red{GOOD} design we'd suggest is
to detect the use of lock (or similar) emojis in the title and replace
them with unambiguous text or symbols.  Unfortunately, none of the
subject apps has a similar design (not even the CCT/SF
implementations).
Therefore, we only consider their own \name implementations not showing the
title as GOOD designs.  Except for those CCT apps which choose not to
show the title, only \textit{Facebook} (case 3 in \myfig~\ref{fig:UI}) and \textit{KakaoTalk}
\red{have} a \red{GOOD} design in this test.

In T7, we consider designs that prioritize the display of domain name
over subdomain name GOOD.  \textit{LINE} is the only GOOD one in our
tests (Case 8).

\textbf{NEUTRAL.} When the lock emoji in the title is shown, we
consider it NEUTRAL if either the full URL or an HTTP indicator (e.g.,
exclamation mark icon) is also shown, so that end users still have a
way of telling that the page is insecure.  Examples include CCT
implementation, \textit{LINE} (Case 4), and \textit{Twitter}.

In evaluating T7, we find that 9 of 18 apps with \red{their} own \name
implementations only display the title of the page without the URL
(e.g., \textit{WeChat} in Case 9), which strictly speaking does not
fall short on the extended subdomain but is still misleading. We
therefore categorize this as NEUTRAL.

\textbf{BAD.} In T6, 11 apps display the lock emoji without any HTTP
indicator or displaying the complete URL (e.g., Cases~5 and~6).  End
users have a high chance \red{of} being misled by the lock emoji, and we
consider such designs BAD.  Regarding T7, 8 of the subject apps
display the subdomain name with the domain name missing (Case 10).

\textbf{iOS.} Again, most apps have similar behavior on their Android
and iOS version, with the exception of \textit{FB Messenger} iOS app
which only has the domain name on its address bar and is considered as
a GOOD design for T6.  For T7, \textit{FB Messenger} iOS app displays
both the head and the tail of the domain name, which is also a \red{GOOD}
design.
\textit{LinkedIn} iOS app, on the other hand, suffers with
displaying the subdomain (BAD) although its Android counterpart only
shows the title (NEUTRAL).

\begin{center}
  \shadowbox{\parbox{0.45\textwidth}{
\textbf{\small Takeaways in \S\ref{RQ2:Page_Displaying}: 
	\begin{itemize}
	    \item Ten apps using Chrome Custom Tabs or SF perform GOOD
              on nearly all the tests from T2 to T7.
	    \item More than half of the remaining 15 apps do not
              display the domain name in their own \name
              implementation, and nearly none of them provides HTTP
              and HTTPS indicators.  This makes it difficult for end
              users to differentiate \red{between} secure and insecure pages.
              Fortunately, all apps behave GOOD when handling URLs
              with certificate errors.
        \item Moreover, nearly all those 15 apps have \red{BAD} performance on handling a 
              title with the fake lock emoji or a long subdomain name, which
              could lead to insecure pages being misinterpreted as
              secure ones.\normalsize
	\end{itemize}
}}}
\end{center}

\subsection{Usability Risks on Page Navigation}
\label{RQ3:Page_Navigating}

\begin{figure}[t!]
    \includegraphics[width=0.4\textwidth]{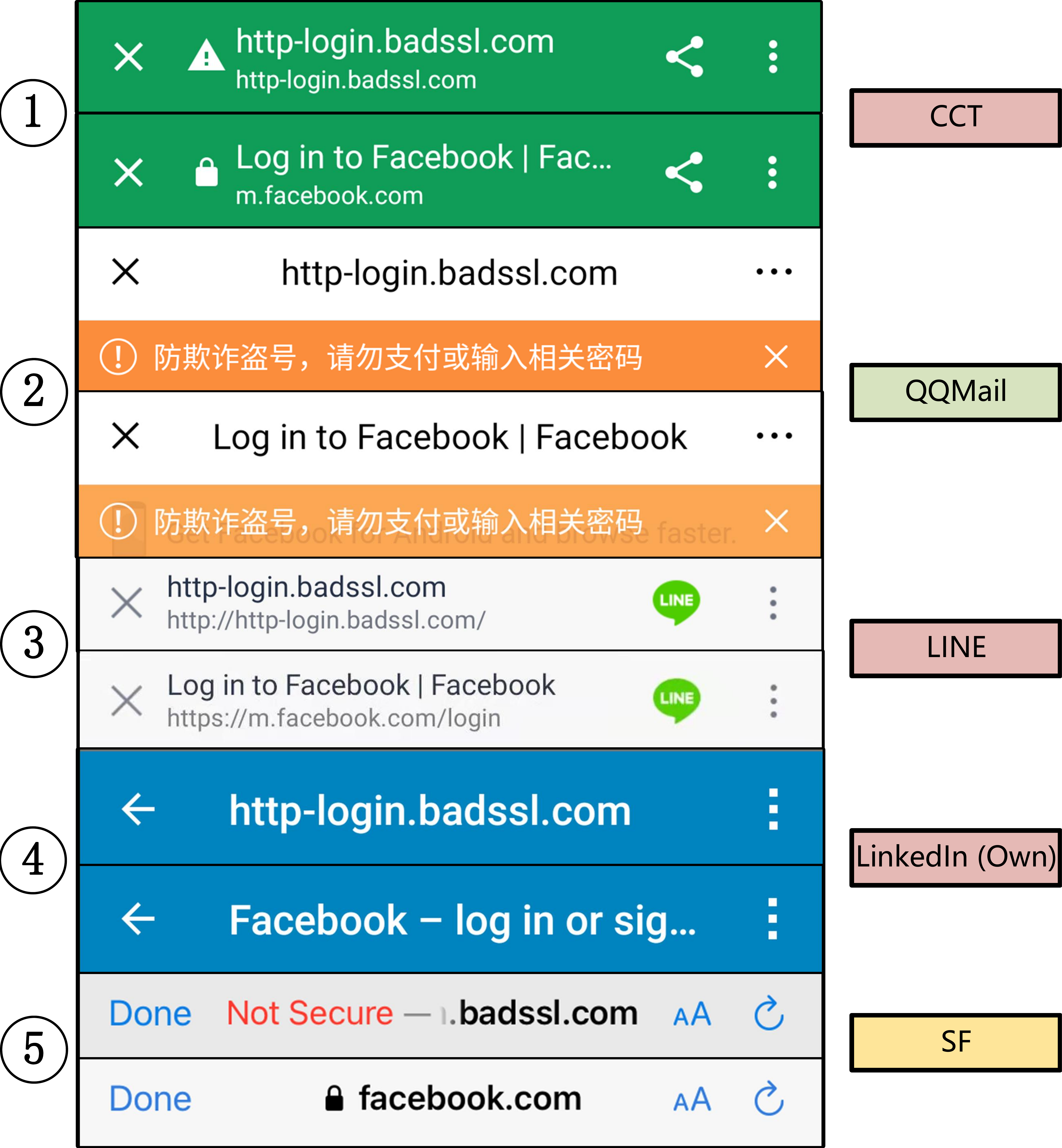}
    \caption{Examples of \red{whether} displaying \red{specific} warnings in the title bar when users browse \red{a (potentially phishing)} login page (T8). \red{We tested both HTTP and HTTPS login pages.}}
    \label{fig:Login}
\end{figure}

\red{As mentioned in Section~\ref{sec:pageNavigate}, our last test (T8) is to see how a subject app's \name implementation reacts to dangerous operation (e.g., password inputting) during the navigation of a web page.
Specifically, T8 tests whether \names would show a \textit{specific} or \textit{extra} warning on a login page as compared with their normal behaviors on a non-login page.
}
\myfig~\ref{fig:Login} shows \red{the} screenshots of some representative \red{\names}.

\textbf{Chrome Custom Tabs and SFSafariViewController.}
\red{As sho-
wn in Case~1 of Figure~\ref{fig:Login},} CCT does not provide any \red{additional warning for password inputting when navigating both HTTP and HTTPS login pages.}
\red{That said, it has the same behaviors as previously in T3 and T4; see \myfig~\ref{fig:HTTPS} and \ref{fig:HTTP}.
Therefore, we consider CCT's rating as BAD in this test.
In contrast, SF performs slightly better than CCT.
As shown in Case~5 of Figure~\ref{fig:Login}, SF highlights the display of ``Not Secure'' (normally displayed for all HTTP pages; see Case 8 in \myfig~\ref{fig:HTTP}) by changing its color to red when end users navigate an HTTP login page,} which is more striking for end users to notice. 
\red{However, SF does not have specific warnings on HTTPS login pages.
As a result, we assign NEUTRAL to SF's performance on T8.}

\textbf{GOOD.} \red{According to our specification in Section~\ref{sec:pageNavigate}, we consider those that provide specific warnings for both HTTP and HTTPS login pages as GOOD designs. }
\textit{QQ Mail} (Case 2) and \textit{QQ} (both of which are developed by Tencent) display such prompt when \red{the} user enters username or password into both HTTP (URL8) and HTTPS (URL10) web page, which are GOOD.

\textbf{NEUTRAL.} 
\red{Some apps display an extra \red{warning} (other than the normal HTTP indicator) \red{during navigation of an} HTTP \red{login} page but fail to provide any \red{warning} on an HTTPS \red{login} page.
According to our specification in Section~\ref{sec:pageNavigate}, we rate these apps' \names as NEUTRAL.
For all tested apps, we find that only the apps using SF have such performance.}

\textbf{BAD.} The other apps score BAD in T8, including
\textit{WeChat} which is also developed by Tencent but they
do not show any \red{warnings} in this test.
Upon further investigation, we find that this anti-fraud tip which is used by the other two apps can be removed in \textit{WeChat} if the domain name is registered on the developer platform of \textit{WeChat}. So maybe some developers have registered this domain name on that platform so the prompt is removed.

\textbf{iOS.} iOS apps their with own \name implementation behave the same as their Android versions in this test.


		

\begin{center}
  \shadowbox{\parbox{0.45\textwidth}{
\textbf{\small \red{Takeaway in \S\ref{RQ3:Page_Navigating}:
    Most of the tested \names do not provide specific warnings to remind users of the risk of inputting passwords during navigating a login page, regardless it uses HTTP or HTTPS. Even CCT/SF does not perform well in this teszhoumot.\normalsize}}
}}
\end{center}

\section{App Developers' Responses}
\label{sec:response}

To understand developers' reaction on our findings and to potentially provide our recommendations on fixing severe \name issues, we issued security reports to all affected apps including \textit{WeChat, FB Messenger, QQ, Snapchat, Instagram} (fixed as we reported), \textit{Weibo, 163 Mail, QQ Mail, Alipay, LinkedIn, Toutiao}, and \textit{Baidu}, through their bug bounty programs or security contact emails\footnote{Apps using CCT/SF are generally secure; so we skip them.
We failed to locate any feedback channel for \textit{KakaoTalk} and \textit{Zhihu} and therefore have to skip them as well.}.
Most of the apps acknowledged our findings and agreed with our assessment, but refused to recognize them as {\em vulnerabilities}, i.e., they consider the reported issues out of the scope of their bug bounty programs.
By analyzing their responses in detail as follows, we find that developers' willingness and readiness to fix usability security issues are rather low compared to fixing technical vulnerabilities, which is a puzzle in usability security research.

\textbf{Facebook's response.}
While the \textit{Facebook} app performs well in nearly all the tests, the other two apps from the same company, namely \textit{FB Messenger} and \textit{Instagram}, did not use the same \name design and failed in our tests T3$\sim$T4 and T6$\sim$T8 (see \mysec\ref{sec:evaluate}).
When we prepared the security reports for these two apps on 4 February 2021, we found that the latest version of \textit{Instagram} had changed its \name design to display a lock icon and an exclamation mark to indicate the HTTPS and HTTP pages, respectively.
This suggests that they also noticed this problem (before our reporting) and made an improvement.
Therefore, we focus on the response to our reports to \textit{FB Messenger}.

There are two key points in the response given by the Facebook security team.
First, they appreciated our report but said that our report does not qualify for their bug bounty program due to the social engineering nature of our reported attacks.
Second, they already have a URL detection system called Linkshim, which could detect potentially malicious URLs and thus defend against \name attacks.
While we agree that the reported \name usability issues would not cause the same security consequences as in the technical WebView vulnerabilities (e.g.,~\cite{WebViewAtk11, FileCross14, MoST15, CrossAppWebViewInfection17, yang2019iframes}), the usability weaknesses in \textit{FB Messenger}'s \names we demonstrated in \mysec\ref{sec:evaluate} definitely give attackers ample rooms to successfully launch phishing attacks.
Moreover, it is a puzzle to see that different apps developed by the same company make vastly different security decisions for seemingly the same component.

\textbf{Snapchat's response.}
Compared with \textit{Facebook}, the security team of \textit{Snapchat} is more concerned on \names' usability security issues with the following response (despite that a priority fix would not be issued as similar to \textit{FB Messenger}):
``\textit{While this attack scenario is quite interesting, we would consider this to be more of a defense-in-depth issue, rather than a discrete security vulnerability in Snapchat itself. Per our program rules, we generally don't accept issues pertaining to `Reports solely indicating a lack of a possible security defense'. While we appreciate your suggestions here, we don't feel that this poses a severe enough risk to warrant a priority fix, and as such we'll be closing this report as Informative. We do appreciate your efforts here, and we hope you'll continue reporting security issues to us in the future.}''

\textbf{LinkedIn's response.}
Compared to \textit{Facebook} and \textit{Snapchat}, the response from \textit{LinkedIn}'s security team is more positive.
Specifically, they consider patching it in the future versions of the \textit{LinkedIn} app:
``\textit{Thank you for your detailed email and report, we appreciate it. We regularly review incoming reports to identify opportunities to improve our member experience and their safe interactions on the platform. We have taken note of these items and included it for consideration in our future roadmaps. Once again, we appreciate the detail and effort that went into this research \& report.}''

\textbf{Some Chinese IT companies' responses.}
We also received responses from a few large Chinese IT companies, including Tencent (developing \textit{QQ, WeChat}, and \textit{QQ Mail} apps), Alibaba (developing \textit{Alipay}), ByteDance (developing \textit{Toutiao}), Sina (developing \textit{Weibo}), NetEase (developing \textit{163 Mail}), and \textit{Baidu}.
Most of them did not take our security reports seriously --- either indicating that it was a known problem (by Tencent) or simply closed our reports without an explanation (by Sina).
The only exception goes to ByteDance, whose security team responded that they just followed ``the industry's standard'' but will report this issue to their product team.

All these responses suggest a play down from app developers when it
comes to \name usability security concerns.

\section{Secure \name Design Principles}
\label{sec:goodDesign}

In this section, we propose a set of secure \name design principles and corresponding code-level implementations in this section to
help mitigate risky \names and guide future designs. \red{Here we provide implemenations in Android as examples, and iOS developers can use corresponding counterparts in iOS.}

Firstly, we recommend the use of Chrome Custom Tabs and
SFSafariViewController for their good performance in our tests presented in this paper \red{except for T8}. The implementation guides are CCT~\cite{CCTGuide} and SF~\cite{SFSafariViewController}. 
Comparing to building one's own IABI, CCT/SF are easy to incorporate with little effort while achieving outstanding security design and optimized loading speed. 

\red{However, CCT and SF also have their limitations}.
\red{CCT failed to provide an extra prompt to alert users when entering passwords on the web page. And they can only provide some basic customization options while developers may need deeper customization to fit in with their apps. In some region, Chrome (and therefore CCT) is not available, e.g., in the mainland China market.}

\red{Considering these shortcomings of CCT/SF, we propose the following \name design principles for those developers who need their own \name implementation. We devide them into three parts: design principles before opening the URL, on page displaysing and on page navigating.}
\red{Considering these shortcomings of CCT/SF, we propose the following \name design principles for developers who need their own \name implemenation. We devide them into three parts: design principles before opening the URL, on page displaysing, and on page navigating.}
\begin{enumerate}[leftmargin=1em, itemsep=0pt, label=\textbf{\arabic*.}]
  \item \textbf{Before Opening the URL:} In the chatting, posting, email UI and other possible UI that display the clickable URL, \names:
  \begin{enumerate}[leftmargin=2em, itemsep=0pt, label=\alph*)]
    \item  \textbf{\textit{SHOULD display the complete URL and corresponding indicators of URL schemes.}}
    It would be GOOD that the indicator be more intuitive and eye-catching than title and favicon.
    \item \textbf{\textit{SHOULD NOT display any extra pre-loading information}} (e.g., favicion and title), unless that URL can be trusted.
  \end{enumerate}

  \item \textbf{On Page Displaying.} After the user taps the URL, developers could adopt the following five principles to better
  \red{avoid the potential usability security issues on page displaying}:

  \begin{enumerate}[leftmargin=2em, itemsep=0pt, label=\textbf{\alph*)} ]
    \item \textbf{\textit{SHOULD display the full URL in the address bar}} to show the page origin.
      Developers can get the URL of the current page through \texttt{WebView.getURL()} or the arguments in the event handlers of \texttt{WebViewClient} (e.g., \texttt{onPageFinished}, shown in Listing~\ref{Code:DisplayURL}).

      \item \textbf{\textit{SHOULD display the HTTP and HTTPS indicators}}, which are intuitive for users to identify insecure web pages, potentially in conjunction with the scheme text in the URL.
      To this end, developers can override the \texttt{onPageFinished} method~\cite{onpagefinished}. Here we provide a simple example in Listing~\ref{Code:DisplayURL}.

\begin{lstlisting}[caption={Display the URL and indicators.},label = {Code:DisplayURL}]
public void onPageFinished(WebView view, String url){
  ...
  //Display url on the title bar.
  addressBar.setText(url);
  //obtain the scheme
  String scheme = URL(url).getProtocol();
  if(scheme.equals("https")){
    /*Display the HTTPS indicator*/
  }else{
    /*Display the insecure indicator*/
  }
  ...
}
\end{lstlisting}


      \item \textbf{\textit{SHOULD NOT directly open URLs with certificate errors.}}
          \begin{enumerate}[leftmargin=2em, itemsep=0pt, label=\alph*.]
              \item Show a prompt, like a dialog box or a special page,
                which informs end users about SSL errors.
              \item Provide end users with the option to continue
                opening the URL in a covert manner, e.g., as in CCT
                which only shows the continue option after clicking on
                the ``Advanced button'' (Case~2 in
                Figure~\ref{fig:Expired}).
          \end{enumerate}
      To handle the certificate errors in the WebView, developers can override the event handler \texttt{WebViewClient.onReceiv-
      edSslError}~\cite{onReceivedSslError}, and show a dialog to inform the user about the error. 

      \item \textbf{\textit{SHOULD handle the lock emoji in the title with extra care}} by:
          \begin{enumerate}[leftmargin=2em, itemsep=0pt, label=\alph*.]
              \item Replacing it with the text to avoid misinterpreting as the HTTPS indicator; or
              \item Disallowing emoji; or
              \item Avoiding displaying the title.
          \end{enumerate}
      Developers can override the \texttt{onPageFinished} method, obtain the title of the web page by \texttt{WebView.getTitle()}, and detect the emoji code in the title. For example, the Unicode of the lock emoji is \textit{U+1F512}. Another choice is to disallow all the unicode in the title. We do not recommend it as it will greatly damage the user experience.   
      
      \item \textbf{\textit{SHOULD handle the long subdomain name with extra care.}} by:
      \begin{enumerate}[leftmargin=2em, itemsep=0pt, label=\alph*.]
        \item Providing scrolling capability for end uses to read the complete domain name; or
        \item Prioritizing the display of domain name over subdomain name.
      \end{enumerate}
      To scroll the \texttt{TextView} (displaying the URL/domain name)in Android apps, developers can set its attribute in the layout xml file: \quad \texttt{android:ellipsize=``marquee''}.  \\
      To Prioritize the domain name, developers can set the attribute to: \quad \texttt{android:ellipsize=``start''}.
  \end{enumerate}

  \item \textbf{On Page Navigating.} When the user tries to enter the password or other sensitive information in an web page, \names:
    \begin{enumerate}[leftmargin=2em, itemsep=0pt, label=\textbf{\alph*)}] 
      \item \textbf{\textit{SHOULD show an additional warning regardless of HTTPS or HTTP pages.}} Case 2 in Figure~\ref{fig:Login} is a good example. \\To detect the input box of the password and username, developers can utilize the interaction between the Java and JavaScript code, i.e., using \texttt{WebView.loadUrl()} to execute JavaScript code to detect the `password' type within the web page and give a corresponding prompt. An example is shown in Listing~\ref{Code:prompt}.
    \end{enumerate}
\begin{lstlisting}[caption={Display a prompt when entering passwords.},label = {Code:prompt}]
webView.loadUrl(
"javascript:(function(){" +
	"var objs=document.getElementsByTagName(\"input\");" 
+ "for(var i=0;i<objs.length;i++)" +"{"
  + "var type = objs[i].getAttribute(\"type\");"
    + "if(type==\"password\"){"  
      + "objs[i].onfocus=function(){"
         + "PROMPT_TO_ENTERING_PASSWORD" 
       +"}" 
    + "}" 
   +"}" 
 + "})()");
\end{lstlisting}
\end{enumerate}


\section{Discussion}
\label{sec:discussion}

\red{In this section, we discuss threats to the validity of our study and limitations.
Specifically, the major threats are that we did not conduct a user study to evaluate \names and our ratings are subjective assessments solely based on designs of the apps' user interfaces and corresponding logic.
Additionally, we discuss our limitations on the lack of a large dataset, automatic testing, and evaluation of the \name design principles.}

\red{\textbf{User study.}
The usability problems mentioned in this paper are not verified in a study with end users, resulting in the lack of direct confirmation of our findings.  For example, in T6, we did not test whether an end user is actually misled by the fake lock emoji in the title, even though an expert analysis on the app's user interface strongly suggests the possibility.
A possible extension of this work is to design an \name app practicing GOOD \name principles and compare end user reactions with those from popular apps tested in this paper.}

\red{\textbf{Ratings.}
  Our evaluations, in particular, the setting of various ratings, are based on previous work~\cite{15TMC_indicator, TMC13SSL, ICS12measuring} and World Wide Web (W3C) guidelines on mobile browsers~\cite{W3C}.  We believe that some of these settings are subjective assessments and do not advocate the uniqueness of such settings.}

\red{\textbf{Dataset.}
We conducted tests only on a relatively small dataset with 25 mobile apps.
However, these apps are all the most popular ones containing \names and are used in daily life.
Therefore, we believe that they capture \names' representative behaviors experienced by end users.
It is worth noting that we do not consider the apps that do not have \names, such as \textit{Whatsapp} and \textit{Signal}, because they jump to default browsers when opening a web page.
}

\red{\textbf{Towards automatic testing.}
Our manual testing limits the scalability of this paper.  Here we explain briefly why an automatic testing is non-trivial with either dynamic or static approaches.}

\red{Unlike existing work on generic dynamic analysis of mobile apps, our analysis of \name behavior requires triggering {\em specific} behavior of mobile apps.  This typically mandates signing up an account with each app and triggering the processing of specific URLs, which is, to say the least, non-trivial.  For example, we cannot find a unified (dynamic) way of precisely locating the (all) chat UI of each app.}

\red{Static analysis also encounters specific challenges, e.g., locating the HTTP and HTTPS indicators, which typically do not present themselves as layout files but images.  Therefore, it is non-trivial to perform backward tracing to find the relationship between these indicators and the web pages.}

\red{\textbf{Evaluating design principles.}
  The principles in Section~\ref{sec:goodDesign} present lessons learned from our systematic analysis of the 25 high-profile apps, but they have not been tested in real-world app development or end-product user studies.  We hope that our study can bring attention to the community and promote more research on the principles and guidelines for \names development.}


\section{Related Work}
\label{sec:related}



\red{In this section, we review some closely related works on the security indicators in regular mobile browsers, the general mobile WebView and TLS security.}

\textbf{Security Indicators in Mobile Browsers.}
Amrutkar et al.~\cite{15TMC_indicator} first measured the adequacy of critical security indicators in mobile browsers. 
They found that mobile browser's UI designs failed to meet many security guidelines.
Luo et al.~\cite{CCS17hindsight} revealed a number of UI vulnerabilities among mobile browsers, which attackers can use to better social engineer users and collect sensitive information.
Wu et al.~\cite{SIGCHI06} evaluated the usability of mobile browser\red{s'} address bar\red{s} for security guarantees.
Different from these studies on standalone mobile browser apps, our study is the first one targeting security indicators in in-app browsing interfaces (\names).
Moreover, we show that the problems become worse when it comes to \names, as many developers only care about the main functionality of the app without putting too much effort into the design of browsing interfaces.

\textbf{Mobile WebView Security.}
Android WebView had been vulnerable to various attacks.
Luo et al. performed the first study on the attacks of WebView~\cite{WebViewAtk11,Touchjacking}, followed by the file-based cross-zone scripting attack~\cite{FileCross14} and access control problem by Georgiev et al.~\cite{NoFrak14}.
Wu and Chang~\cite{MoST15} further \red{studied} the WebView vulnerabilities on the iOS platform.
There are also many techniques to prevent private data from leaking through JavaScript, for example, BavelView~\cite{BabelView18}, Spartan Jester~\cite{SpartanJester17}, and HybriDroid~\cite{HybriDroid16}.
Most of the past research focused on the interaction between Java and JavaScript but not on the usability security of the in-app browsing interfaces.
For example, Li et al.~\cite{CCS17Unleashing} proposed attacks that utilize browsing interfaces for cross-app navigation from another WebView.
Similarly, Yang et al.~\cite{yang2019iframes} found that iframe can navigate the WebView to untrusted web pages.

\textbf{Mobile App TLS Security.}
Many Android apps use SSL/TLS to transmit sensitive information securely, but developers often use their own, potentially insure, implementation to verify the certificate. Georiev et al. showed that SSL certificate validation is completely broken in various popular apps and libraries~\cite{CCS12validateSSL}. Thus, many previous works studied the potential security threats caused by the inadequate or insecure use of TLS in mobile browsers. They provided various tools to detect potential vulnerabilities against Man-In-The-Middle attacks caused by inadequate use of SSL/TLS, \red{including the dynamic} MalloDroid~\cite{SSLbyMalloDroid12} and SMV-Hunter~\cite{SMVHunter14} \red{and the static Amandroid~\cite{Amandroid14} and BackDroid~\cite{BackDroid21}}.
In this paper, we find that most of the apps that use their own \names do not have any security indicators about the schemes the website is using, to the extent that the users cannot identify whether the current web page they are browsing meet the TLS requirement.

\section{Conclusion}
\label{sec:conclude}

In this paper, we conducted the first empirical study on the usability (in)security of in-app browsing interfaces (\names) in both Android and iOS apps.
Atop a dataset of 25 high-profile mobile apps that contain \names, we performed a systematic analysis that comprises eight security tests and covers all the attack surfaces from opening, displaying, to navigating an in-app web page.
We obtained three major security findings, including about 30\% of the tested apps fail to provide enough URL information before users open the URL, nearly all custom \names have various problems in providing sufficient indicators to faithfully display an in-app page to users, and only a few \names give \red{specific warnings} to remind users \red{of} dangerous operations \red{(e.g., password inputting)} during \red{navigating a login} page.
To help mitigate risky \names and guide future designs, we reported our findings to affected vendors, analyzed their \red{responses}, and proposed a set of secure \name design principles.

\begin{acks}
\red{We thank our shepherd, Yasemin Acar, for her comprehensive guidance and the anonymous reviewers for their valuable comments and suggestions. This research/project is partially supported by the Singapore National Research Foundation under the National Satellite of Excellence in Mobile Systems Security and Cloud Security (NRF2018NCR-NSOE004-0001) and a direct grant (ref. no. 4055127) from The Chinese University of Hong Kong.}
\end{acks}

\balance
\bibliographystyle{ACM-Reference-Format}
\bibliography{main}

\end{document}